\documentclass[10pt]{article}
\usepackage{graphicx}
\usepackage{amsmath}
\usepackage{amssymb}
\usepackage{caption2}
\begin{document}
\bibliographystyle{prsty}
\begin{center}
{\large {\bf \sc{  Analysis of  the $P_{cs}(4338)$ and related pentaquark molecular states  via the QCD sum rules }}} \\[2mm]
 Xiu-Wu Wang,   Zhi-Gang Wang \footnote{E-mail: zgwang@aliyun.com.  }      \\
 Department of Physics, North China Electric Power University, Baoding 071003, P. R. China
\end{center}

\begin{abstract}
In this work, we tentatively identify the $P_{cs}(4338)$ as the $\bar{D}\Xi_c$  molecular state, and distinguish  the isospins of the current operators to  explore  the $\bar{D}\Xi_c$,  $\bar{D}\Lambda_c$, $\bar{D}_s\Xi_c$, $\bar{D}_s\Lambda_c$, $\bar{D}^*\Xi_c$,  $\bar{D}^*\Lambda_c$, $\bar{D}^*_s\Xi_c$ and $\bar{D}^*_s\Lambda_c$   molecular states  without strange, with strange  and with double strange  in the framework of  the QCD sum rules in details.  The present explorations   favor  identifying the $P_{cs}(4338)$ ($P_{cs}(4459)$) as  the $\bar{D}\Xi_c$ ($\bar{D}^*\Xi_c$)  molecular state with the spin-parity  $J^P={\frac{1}{2}}^-$ (${\frac{3}{2}}^-$) and isospin  $(I,I_3)=(0,0)$, the observation of their cousins with the isospin $(I,I_3)=(1,0)$ in the $J/\psi\Sigma^0/\eta_c\Sigma^0$ invariant mass distributions   would
decipher  their  inner structures.
\end{abstract}

 PACS number: 12.39.Mk, 14.20.Lq, 12.38.Lg

Key words: Pentaquark molecular states, QCD sum rules

\section{Introduction}

In 2015,  the  LHCb collaboration observed  the $P_c(4380)$ and $P_c(4450)$ in the $J/\psi p$ invariant  mass distribution  with the  favored  spin-parity  $J^P={\frac{3}{2}}^-$ and  ${\frac{5}{2}}^+$ (rather than the reversed parity ${\frac{3}{2}}^+$ and  ${\frac{5}{2}}^-$) respectively in the $\Lambda_b^0\to J/\psi K^- p$ decays \cite{LHCb-4380}. If they are really resonant states (not re-scattering effects, threshold effects, cusp effects),    they must have minimal valence quark content of $uudc\bar{c}$ and are excellent pentaquark candidates.
In 2019,  the LHCb collaboration confirmed the structure $P_c(4450)$, which is resolved with two narrow overlapping peaks $P_c(4440)$ and $P_c(4457)$ with the statistical significance of $5.4\,\sigma$.  In addition, they   observed a  narrow structure $P_c(4312)$ in the $ J/\psi  p$ invariant mass distribution with the
statistical significance of $7.3\,\sigma$ in a much larger data sample, the $P_c(4312)$ is also an excellent pentaquark candidate with the minimal valence quark content $uudc\bar{c}$ \cite{LHCb-Pc4312}.

 In 2020, the LHCb collaboration reported an evidence of new structure $P_{cs}(4459)$ in the $J/\psi \Lambda$ invariant mass distribution  with  a significance of  $3.1\sigma$ in the $\Xi_b^- \to J/\psi K^- \Lambda$ decays \cite{LHCb-Pcs4459-2012}. If the $P_{cs}(4459)$ is confirmed to be a real resonance, it is an excellent  pentaquark candidate with the minimal valence quark content $udsc\bar{c}$. In fact, the $P_{cs}(4459)$
 is also consistent with being due to two resonances, just like in the case of the $P_c(4450)$.

 In 2021,  the LHCb collaboration observed an evidence for a new structure $P_c(4337)$ in the $J/\psi p$ and $J/\psi \bar{p}$ systems in the $B_s^0 \to J/\psi p \bar{p}$ decays with  a significance in the range of $3.1-3.7\sigma$ depending on the assigned $J^P$ hypothesis \cite{LHCb-Pc4337-2108}. The existence of the $P_c(4337)$ is still need confirmation and its spin-parity are still need measurement.

 Recently, the LHCb collaboration observed an evidence for a new structure $P_{cs}(4338)$ in the $J/\psi \Lambda$ mass distribution in the $B^- \to J/\psi \Lambda \bar{p}$ decays \cite{LHCb-Pcs4338}. The measured  Breit-Wigner mass and width are $4338.2\pm0.7\pm0.4\,\rm{MeV}$ and $7.0\pm1.2\pm1.3\,\rm{MeV}$ respectively and the favored  spin-parity is $J^P={\frac{1}{2}}^-$. The $P_{cs}(4338)$ and $P_{cs}(4459)$ are observed in the $J/\psi \Lambda$ invariant mass distribution,  they have the isospin $I=0$, as the strong decays conserve the isospin, the observation of their isospin cousins are of crucial importance.

 The  $P_c(4312)$, $P_c(4380)$, $P_c(4440)$, $P_c(4457)$, $P_{cs}(4459)$ and $P_{cs}(4338)$ lie slightly below or above the thresholds of the charmed  meson-baryon pairs $\bar{D}\Sigma_c$, $\bar{D}\Sigma_c^*$, $\bar{D}^*\Sigma_c$, $\bar{D}^*\Sigma_c$, $\bar{D}\Xi_c^\prime$ ($\bar{D}\Xi_c^*$, $\bar{D}^*\Xi_c$, $\bar{D}^*\Xi_c^\prime$) and $\bar{D}\Xi_c$, respectively. Just as what we expected, the  $P_c(4312)$, $P_c(4380)$, $P_c(4440)$, $P_c(4457)$, $P_{cs}(4459)$ and $P_{cs}(4338)$ have been tentatively assigned to be the charmed meson-baryon molecular states according to several phenomenological analysis    \cite{Pcs4459-mole-BSE-CC,Penta-mole-CREFT,Penta-mole-OBE-CC,Penta-mole-BSE-CC,Pcs4459-mole-BSE-2-CC,
 Penta-mole-ERE-CC,Penta-mole-WangB}. While it is difficult to identify  the $P_c(4337)$ as the  molecular state without resorting to the help of  large coupled-channel effects  due to lacking nearby  meson-baryon thresholds,  it is more natural to identify  the $P_c(4337)$ as the $A-A-\bar{c}$-type hidden-charm pentaquark state with the spin-parity $J^P={\frac{1}{2}}^-$, where the $A$ denotes the axialvector diquark states \cite{WZG-IJMPA-penta}.

In fact, we can reproduce the masses of the $P_c(4312)$, $P_c(4337)$, $P_c(4380)$, $P_c(4440)$, $P_c(4457)$ and $P_{cs}(4459)$ in the picture  of  diquark-diquark-antiquark type (or diquark-triquark type) pentaquark states  via the theoretical method of QCD sum rules \cite{WZG-IJMPA-penta,WZG-Pcs4459}.

 In Ref.\cite{WZG-WHJ-XQ}, we adopt  the  hadronic dressing  mechanism  to compromise the pentaquark and  molecule interpretations, which are two quite different schemes but both can give satisfactory experimental  masses.  The hadronic dressing  mechanism was introduced previously to interpret the exotic  hierarchy of the masses of the scalar mesons below $1 \,\rm{GeV}$  \cite{Hadron-dress-1,Hadron-dress-2,Hadron-dress-3}, and we expect the same mechanism exists  here.   The pentaquark states maybe have a diquark-diquark-antiquark type intrinsic pentaquark
kernel $qqqc\bar{c}$ with the typical size of the conventional  $qqq$  baryons, say about $0.5\sim 0.7 \,\rm{fm}$,  the strong couplings of the intrinsic  kernels  $qqqc\bar{c}$  to the nearby charmed  meson-baryon pairs result in  some  molecule  components, and  the pentaquark states maybe spend  a rather  large time as the  molecular states, thus they maybe  display properties of the molecules.

According to our previous  calculations in the framework of the QCD sum rules, the lowest diquark-diquark-antiquark type hidden-charm pentaquark state without strange  has the mass about $4.31\,\rm{GeV}$, it is difficult to assign the $P_{cs}(4338)$ as the  diquark-diquark-antiquark type  pentaquark state with strange, or there exists contradiction in identifying the $P_c(4312)$ and $P_{cs}(4338)$ in the same picture as pentaquark states. The lowest diquark-diquark-antiquark type hidden-charm pentaquark state with strange has a mass about $4.45\,\rm{GeV}$ according to the direct calculations in the framework of the QCD sum rules and qualitative  analysis based on the $SU(3)$ mass-breaking effects  \cite{WZG-IJMPA-penta,WZG-Pcs4459}, which is much larger than the mass of the $P_{cs}(4338)$.

In the QCD sum rules, we usually choose the local currents to interpolate the tetraquark or pentaquark  molecular states  which have  two color-neutral clusters \cite{Wang-molecule-QCDSR-2,WZG-hidden-charm-mole,Penta-mole-QCDSR-Chen,Penta-mole-QCDSR-Zhang,Penta-mole-QCDSR-Azizi,WZG-IJMPA-mole,Pcs4459-mole-QCDSR}, the color-neutral clusters  are not necessary to be the physical mesons and baryons, they just have the same quantum numbers as the physical mesons and baryons. The local currents require that the  molecular states have the average spatial sizes $\sqrt{\langle r^2\rangle}$ of the same magnitudes as the conventional mesons and baryons, and they are also compact objects, just like the diquark-antidiquark type
tetraquark states or diquark-diquark-antiquark type pentaquark states, and those molecular states are not necessary to be loosely bound, as the conventional mesons and baryons  are  compact objects, in the local limit, the conventional  mesons and baryons  lose themselves and
merge into color-singlet-color-singlet type tetraquark or pentaquark states \cite{WZG-color-neutral}.

  In the present work, we extend our previous works on the pentaquark molecular states to investigate  the  $\bar{D}\Xi_c$,  $\bar{D}\Lambda_c$, $\bar{D}_s\Xi_c$, $\bar{D}_s\Lambda_c$, $\bar{D}^*\Xi_c$,  $\bar{D}^*\Lambda_c$, $\bar{D}^*_s\Xi_c$ and $\bar{D}^*_s\Lambda_c$   molecular states   with distinguished isospins in the framework of  the QCD sum rules \cite{WZG-IJMPA-mole,WZG-Pcs-mole,WZG-WXW-IsoSpin,WZG-WZW-Pcs-mole}. We carry out the operator product expansion up to the vacuum condensates  of dimension $13$ consistently, just like what we did previously, and determine the best energy scales of the spectral densities using the modified energy-scale formula by considering the light-flavor $SU(3)$ mass-breaking effects,    and try to obtain the lowest color-singlet-color-singlet type pentaquark states as one of the color-neutral clusters has the same quantum numbers as the lowest charmed baryons in the flavor anti-triplet,   and make possible assignments of the $P_{cs}(4338)$ and $P_{cs}(4459)$.

 The article is arranged as follows:  we acquire the QCD sum rules for the pentaquark  molecular states  in Sect.2;  in Sect.3, we present the numerical results and discussions; and Sect.4 is reserved for our
conclusion.

\section{QCD sum rules for  the  pentaquark molecular states}
Firstly, let us  write down  the two-point correlation functions $\Pi(p)$ and $\Pi_{\mu\nu}(p)$,
\begin{eqnarray}\label{CF-Pi}
\Pi(p)&=&i\int d^4x e^{ip \cdot x} \langle0|T\left\{J(x)\bar{J}(0)\right\}|0\rangle\nonumber \, ,\\
\Pi_{\mu\nu}(p)&=&i\int d^4x e^{ip \cdot x} \langle0|T\left\{J_{\mu}(x)\bar{J}_{\nu}(0)\right\}|0\rangle \, ,
\end{eqnarray}
where the interpolating currents,
 \begin{eqnarray}
 J(x)&=&J_{(0,0)}^{\bar{D}\Xi_c}(x)\, , \, J_{(1,0)}^{\bar{D}\Xi_c}(x)\, , \, J_{(\frac{1}{2},\frac{1}{2})}^{\bar{D}\Lambda_c}(x)\, , \, J_{(\frac{1}{2},\frac{1}{2})}^{\bar{D}_s\Xi_c}(x)\, , \, J_{(0,0)}^{\bar{D}_s\Lambda_c}(x)\, ,\nonumber \\
 J_\mu(x)&=&J_{(0,0)}^{\bar{D}^*\Xi_c}(x)\, , \,J_{(1,0)}^{\bar{D}^*\Xi_c}(x)\, , \,J_{(\frac{1}{2},\frac{1}{2})}^{\bar{D}^*\Lambda_c}(x)\, , \,J_{(\frac{1}{2},\frac{1}{2})}^{\bar{D}_s^*\Xi_c}(x)\, , \, J_{(0,0)}^{\bar{D}_s^*\Lambda_c}(x)\, ,
 \end{eqnarray}
\begin{eqnarray}
J_{(0,0)}^{\bar{D}\Xi_c}(x)&=&\frac{1}{\sqrt{2}}J_{\bar{D}^0}(x)J_{\Xi_c^{0}}(x)-\frac{1}{\sqrt{2}}J_{\bar{D}^-}(x)J_{\Xi_c^{+}}(x) \, , \nonumber\\
J_{(1,0)}^{\bar{D}\Xi_c}(x)&=&\frac{1}{\sqrt{2}}J_{\bar{D}^0}(x)J_{\Xi_c^{0}}(x)+\frac{1}{\sqrt{2}}J_{\bar{D}^-}(x)J_{\Xi_c^{+}}(x) \, , \nonumber\\
J_{(\frac{1}{2},\frac{1}{2})}^{\bar{D}\Lambda_c}(x)&=&J_{\bar{D}^0}(x)J_{\Lambda_c^{+}}(x) \, , \nonumber\\
J_{(\frac{1}{2},\frac{1}{2})}^{\bar{D}_s\Xi_c}(x)&=&J_{\bar{D}^-_s}(x)J_{\Xi_c^{+}}(x) \, , \nonumber\\
J_{(0,0)}^{\bar{D}_s\Lambda_c}(x)&=&J_{\bar{D}^{-}_s}(x)J_{\Lambda_c^{+}}(x) \, ,
\end{eqnarray}
\begin{eqnarray}
J_{(0,0)}^{\bar{D}^*\Xi_c}(x)&=&\frac{1}{\sqrt{2}}J_{\bar{D}^{*0}}(x)J_{\Xi_c^{0}}(x)-\frac{1}{\sqrt{2}}J_{\bar{D}^{*-}}(x)J_{\Xi_c^{+}}(x) \, , \nonumber\\
J_{(1,0)}^{\bar{D}^*\Xi_c}(x)&=&\frac{1}{\sqrt{2}}J_{\bar{D}^{*0}}(x)J_{\Xi_c^{0}}(x)+\frac{1}{\sqrt{2}}J_{\bar{D}^{*-}}(x)J_{\Xi_c^{+}}(x) \, , \nonumber\\
J_{(\frac{1}{2},\frac{1}{2})}^{\bar{D}^*\Lambda_c}(x)&=&J_{\bar{D}^{*0}}(x)J_{\Lambda_c^{+}}(x) \, , \nonumber\\
J_{(\frac{1}{2},\frac{1}{2})}^{\bar{D}^*_s\Xi_c}(x)&=&J_{\bar{D}^{*-}_s}(x)J_{\Xi_c^{+}}(x) \, , \nonumber\\
J_{(0,0)}^{\bar{D}^*_s\Lambda_c}(x)&=&J_{\bar{D}^{*-}_s}(x)J_{\Lambda_c^{+}}(x) \, ,
\end{eqnarray}
and
\begin{eqnarray}
J_{\bar{D}^0}(x)&=&\bar{c}(x)i\gamma_5u(x)\, ,\nonumber \\
J_{\bar{D}^-}(x)&=&\bar{c}(x)i\gamma_5d(x)\, ,\nonumber \\
J_{\bar{D}_s^-}(x)&=&\bar{c}(x)i\gamma_5s(x)\, ,\nonumber \\
J_{\bar{D}^{*0}}(x)&=&\bar{c}(x)\gamma_\mu u(x)\, ,\nonumber \\
J_{\bar{D}^{*-}}(x)&=&\bar{c}(x)\gamma_\mu d(x)\, ,\nonumber \\
J_{\bar{D}_s^{*-}}(x)&=&\bar{c}(x)\gamma_\mu s(x)\, ,\nonumber \\
J_{\Xi_c^{ 0}}(x)&=&\varepsilon^{ijk}d^{T}_i(x)C\gamma_{5}s_j(x) c_k(x)\, ,\nonumber\\
J_{\Xi_c^{ +}}(x)&=&\varepsilon^{ijk}u^{T}_i(x)C\gamma_{5}s_j(x) c_k(x)\, ,\nonumber\\
J_{\Lambda_c^{ +}}(x)&=&\varepsilon^{ijk}u^{T}_i(x)C\gamma_{5}d_j(x) c_k(x)\, ,
\end{eqnarray}
the super(sub)scripts $i, j, k$ are color indices, and the $C$ represents the charge conjugation matrix,
in fact, the $J_{\bar{D}^0}(x)$, $J_{\bar{D}^-}(x)$, $J_{\bar{D}_s^-}(x)$,
$J_{\bar{D}^{*0}}(x)$, $J_{\bar{D}^{*-}}(x)$, $J_{\bar{D}_s^{*-}}(x)$, $J_{\Xi_c^{ 0}}(x)$,
$J_{\Xi_c^{ +}}(x)$ and $J_{\Lambda_c^{ +}}(x)$ are the commonly used meson and baryon currents, respectively,
 the subscripts $(1,0)$, $(0,0)$ and $(\frac{1}{2},\frac{1}{2})$ represent the isospins $(I,I_3)$.

According to quark-hadron duality, the currents $J(0)$ couple potentially to the
  hidden-charm   molecular  states with the spin-parity $J^P={\frac{1}{2}}^\pm$, as for the currents $J_\mu(0)$, they couple potentially to the  hidden-charm   molecular  states with the spin-parity $J^P={\frac{1}{2}}^\pm$ and ${\frac{3}{2}}^\pm$,
\begin{eqnarray}\label{J-lamda-1}
\langle 0| J (0)|P^{-}_{\frac{1}{2}}(p)\rangle &=&\lambda^{-}_{\frac{1}{2}} U^{-}(p,s) \, ,\nonumber  \\
\langle 0| J (0)|P^{+}_{\frac{1}{2}}(p)\rangle &=&\lambda^{+}_{\frac{1}{2}}i\gamma_5 U^{+}(p,s) \, ,
 \end{eqnarray}
 \begin{eqnarray}\label{J-lamda-1X}
\langle 0| J_\mu (0)|P^{-}_{\frac{1}{2}}(p)\rangle &=&f_{\frac{1}{2}}^{-} p_\mu i\gamma_5 U^{-}(p,s) \, ,\nonumber  \\
\langle 0| J_\mu (0)|P^{+}_{\frac{1}{2}}(p)\rangle &=&f_{\frac{1}{2}}^{+}p_\mu U^{+}(p,s) \, ,\nonumber  \\
\langle 0| J_\mu (0)|P^{-}_{\frac{3}{2}}(p)\rangle &=&\lambda_{\frac{3}{2}}^{-}  U^{-}_\mu(p,s) \, ,\nonumber  \\
\langle 0| J_\mu (0)|P^{+}_{\frac{3}{2}}(p)\rangle &=&\lambda_{\frac{3}{2}}^{+}i\gamma_5 U_\mu^{+}(p,s) \, ,
 \end{eqnarray}
where the $\lambda^{\pm}_{\frac{1}{2}}$, $\lambda^{\pm}_{\frac{3}{2}}$ and $f^{\pm}_{\frac{1}{2}}$ are the current-molecule coupling constants (or pole residues), the $U^\pm(p,s)$ and $U_\mu^\pm(p,s)$ are the Dirac spinors and  Rarita-Schwinger spinors, respectively \cite{WZG-IJMPA-penta,WZG-IJMPA-mole,Wang1508-EPJC,WangHuang1508-1}.

 At the hadron side of the correlation functions $\Pi(p)$ and $\Pi_{\mu\nu}(p)$, we isolate the  ground state contributions from the hidden-charm  molecular states with the spin-parity $J^P={\frac{1}{2}}^\pm$ and ${\frac{3}{2}}^\pm$, respectively, and acquire  the hadronic representation \cite{WZG-IJMPA-penta,WZG-IJMPA-mole,Wang1508-EPJC,WangHuang1508-1},
\begin{eqnarray}\label{hadron-rep}
  \Pi(p) & = & \left(\lambda^{-}_{\frac{1}{2}}\right)^2  {\!\not\!{p}+ M_{-} \over M_{-}^{2}-p^{2}  } +  \left(\lambda^{+}_{\frac{1}{2}}\right)^2  {\!\not\!{p}- M_{+} \over M_{+}^{2}-p^{2}  } +\cdots  \, ,\nonumber\\
  &=&\Pi^1_{\frac{1}{2}}(p^2)\!\not\!{p}+\Pi^0_{\frac{1}{2}}(p^2)\, ,\nonumber\\
  \Pi_{\mu\nu}(p) & = & \left(\lambda^{-}_{\frac{3}{2}}\right)^2  {\!\not\!{p}+ M_{-} \over M_{-}^{2}-p^{2}  }(-g_{\mu\nu}) +  \left(\lambda^{+}_{\frac{3}{2}}\right)^2  {\!\not\!{p}- M_{+} \over M_{+}^{2}-p^{2}  }(-g_{\mu\nu}) +\cdots  \, ,\nonumber\\
  &=&-\Pi^1_{\frac{3}{2}}(p^2)\!\not\!{p}g_{\mu\nu}-\Pi^0_{\frac{3}{2}}(p^2)g_{\mu\nu}+\cdots \, ,
\end{eqnarray}
we choose the components $\Pi^{1/0}_{\frac{1}{2}}(p^2)$ and $\Pi^{1/0}_{\frac{3}{2}}(p^2)$ to explore the molecular states with the spin-parity $J^P={\frac{1}{2}}^-$ and ${\frac{3}{2}}^-$, respectively.

 In the following, we omit the subscripts of the pole residues and correlation functions  in above equations, see Eqs.\eqref{J-lamda-1}-\eqref{hadron-rep}, and mark them as the $\lambda_\pm$ and $\Pi^{1/0}(s)$, respectively. It is direct to  get the hadronic  spectral densities  through dispersion relation,
\begin{eqnarray}
\frac{{\rm Im}\Pi^1(s)}{\pi}&=&\lambda^{2}_{-} \delta\left(s-M_{-}^2\right)+\lambda_{+}^2 \delta\left(s-M_{+}^2\right) =\, \rho^1_{H}(s) \, , \\
\frac{{\rm Im}\Pi^0(s)}{\pi}&=&M_{-}\lambda_{-}^2 \delta\left(s-M_{-}^2\right)-M_{+}\lambda_{+}^2 \delta\left(s-M_{+}^2\right)
=\rho^0_{H}(s) \, ,
\end{eqnarray}
where  we introduce the index $H$ to stand for the hadron side,
then we get the QCD sum rules at the hadron side with the help of the  weight functions $\sqrt{s}\exp\left(-\frac{s}{T^2}\right)$ and $\exp\left(-\frac{s}{T^2}\right)$,
\begin{eqnarray}
\int_{4m_c^2}^{s_0}ds \left[\sqrt{s}\rho^1_{H}(s)+\rho^0_{H}(s)\right]\exp\left( -\frac{s}{T^2}\right)
&=&2M_{-}\lambda_{-}^2\exp\left( -\frac{M_{-}^2}{T^2}\right) \, ,
\end{eqnarray}
where the $s_0$ are the continuum threshold parameters and the $T^2$ are the Borel parameters.

It is also direct  to carry out the operator product expansion routinely  \cite{WZG-IJMPA-penta,WZG-IJMPA-mole,Wang1508-EPJC,WangHuang1508-1}.
We contract the $u$, $d$, $s$ and $c$ quark fields in the correlation functions $\Pi(p)$ and $\Pi_{\mu\nu}(p)$  with the Wick's theorem, and observe that there are three full light-quark propagators ($U_{ij}(x)$, $D_{ij}(x)$, $S_{ij}(x)$ in the coordinate space), and two full charm-quark propagators ($C_{ij}(x)$ in the momentum space),
 \begin{eqnarray}
U/D_{ij}(x)&=& \frac{i\delta_{ij}\!\not\!{x}}{ 2\pi^2x^4}-\frac{\delta_{ij}\langle
\bar{q}q\rangle}{12} -\frac{\delta_{ij}x^2\langle \bar{q}g_s\sigma Gq\rangle}{192} -\frac{ig_sG^{a}_{\alpha\beta}t^a_{ij}(\!\not\!{x}
\sigma^{\alpha\beta}+\sigma^{\alpha\beta} \!\not\!{x})}{32\pi^2x^2} -\frac{\delta_{ij}x^4\langle \bar{q}q \rangle\langle g_s^2 GG\rangle}{27648} \nonumber\\
&&  -\frac{1}{8}\langle\bar{q}_j\sigma^{\mu\nu}q_i \rangle \sigma_{\mu\nu}+\cdots \, ,
\end{eqnarray}
\begin{eqnarray}
S_{ij}(x)&=& \frac{i\delta_{ij}\!\not\!{x}}{ 2\pi^2x^4}
-\frac{\delta_{ij}m_s}{4\pi^2x^2}-\frac{\delta_{ij}\langle
\bar{s}s\rangle}{12} +\frac{i\delta_{ij}\!\not\!{x}m_s
\langle\bar{s}s\rangle}{48}-\frac{\delta_{ij}x^2\langle \bar{s}g_s\sigma Gs\rangle}{192}+\frac{i\delta_{ij}x^2\!\not\!{x} m_s\langle \bar{s}g_s\sigma
 Gs\rangle }{1152}\nonumber\\
&& -\frac{ig_s G^{a}_{\alpha\beta}t^a_{ij}(\!\not\!{x}
\sigma^{\alpha\beta}+\sigma^{\alpha\beta} \!\not\!{x})}{32\pi^2x^2} -\frac{\delta_{ij}x^4\langle \bar{s}s \rangle\langle g_s^2 GG\rangle}{27648}-\frac{1}{8}\langle\bar{s}_j\sigma^{\mu\nu}s_i \rangle \sigma_{\mu\nu}  +\cdots \, ,
\end{eqnarray}
\begin{eqnarray}
C_{ij}(x)&=&\frac{i}{(2\pi)^4}\int d^4k e^{-ik \cdot x} \left\{
\frac{\delta_{ij}}{\!\not\!{k}-m_c}
-\frac{g_sG^n_{\alpha\beta}t^n_{ij}}{4}\frac{\sigma^{\alpha\beta}(\!\not\!{k}+m_c)+(\!\not\!{k}+m_c)
\sigma^{\alpha\beta}}{(k^2-m_c^2)^2}\right.\nonumber\\
&&\left. -\frac{g_s^2 (t^at^b)_{ij} G^a_{\alpha\beta}G^b_{\mu\nu}(f^{\alpha\beta\mu\nu}+f^{\alpha\mu\beta\nu}+f^{\alpha\mu\nu\beta}) }{4(k^2-m_c^2)^5}+\cdots\right\} \, ,\nonumber\\
f^{\alpha\beta\mu\nu}&=&(\!\not\!{k}+m_c)\gamma^\alpha(\!\not\!{k}+m_c)\gamma^\beta(\!\not\!{k}+m_c)\gamma^\mu(\!\not\!{k}+m_c)\gamma^\nu(\!\not\!{k}+m_c)\, ,
\end{eqnarray}
and  $t^n=\frac{\lambda^n}{2}$, the $\lambda^n$ is the Gell-Mann matrix
\cite{PRT85,Pascual-1984,WangHuang3900}. If each charm-quark line emits a gluon and each light-quark line contributes  a quark-antiquark   pair, we acquire  a quark-gluon  operator $g_s^2G_{\alpha\beta}G^{\alpha\beta}\bar{q}q \bar{q}q \bar{q}q$ (with $q=u$, $d$ or $s$)  of dimension 13, therefore,  we have to  deal  with  the  condensates at least
up to dimension 13 to judge the convergent behavior of the operator product expansion, as the condensates are vacuum expectations of the quark-gluon operators in the QCD vacuum.

We retain the possible operators $\langle\bar{q}_j\sigma_{\mu\nu}q_i \rangle$ and $\langle\bar{s}_j\sigma_{\mu\nu}s_i \rangle$  from the Fierz transformations of the quark operators
$\langle q_i \bar{q}_j\rangle$ and $\langle s_i \bar{s}_j\rangle$ (before the Wick's contractions) to  absorb the gluons  emitted from other quark lines to  extract the mixed condensates  $\langle\bar{q}g_s\sigma G q\rangle$ and $\langle\bar{s}g_s\sigma G s\rangle$, respectively \cite{WangHuang3900}.  Then we compute  all the integrals in the coordinate space and momentum space sequentially to obtain the representations at the quark-gluon  level.

We count the vacuum condensates  by the strong fine structure constant $\alpha_s=\frac{g_s^2}{4\pi}$ with  the orders $\mathcal{O}( \alpha_s^{k})$, where
$k=0$, $\frac{1}{2}$, $1$, $\frac{3}{2}$, $\cdots$.
In this work, we  prefer  the truncations $k\leq 1$ consistently,
and deal with  the quark-gluon operators of the orders $\mathcal{O}( \alpha_s^{k})$ with $k\leq 1$. To be more precise and concrete, we take account of the vacuum condensates $\langle\bar{q}q\rangle$, $\langle\frac{\alpha_{s}GG}{\pi}\rangle$, $\langle\bar{q}g_{s}\sigma Gq\rangle$, $\langle\bar{q}q\rangle^2$,
$\langle\bar{q}q\rangle \langle\frac{\alpha_{s}GG}{\pi}\rangle$,  $\langle\bar{q}q\rangle  \langle\bar{q}g_{s}\sigma Gq\rangle$, $\langle\bar{q}q\rangle^3$,
$\langle\bar{q}g_{s}\sigma Gq\rangle^2$, $\langle\bar{q}q\rangle^2 \langle\frac{\alpha_{s}GG}{\pi}\rangle$,
  $\langle\bar{q} q\rangle^2\langle\bar{q}g_s\sigma Gq\rangle $,
    $\langle\bar{q} q\rangle \langle\bar{q}g_s\sigma Gq\rangle^2 $,
  $\langle \bar{q}q\rangle^3\langle \frac{\alpha_s}{\pi}GG\rangle$
 with the assumption of vacuum saturation consistently to assess the convergent behaviors  \cite{WZG-Saturation}, where $q=u$, $d$ or $s$.
 In addition, we set the masses of the $u$ and $d$ quarks to be zero and consider  the contributions of the order $\mathcal{O}(m_s)$ consistently for the $s$ quark  so as to take  account of the light-flavor $SU(3)$ mass-breaking effects.

 At last, we acquire the QCD spectral densities $\rho^1_{QCD}(s)$ and $\rho^0_{QCD}(s)$ through   dispersion relation,  their explicit expressions  are available through  contacting the corresponding author via E-mail.
 Then we assume (and implement) the quark-hadron duality below the continuum thresholds  $s_0$, and again we  acquire  the  QCD sum rules  with the help of the  weight functions $\sqrt{s}\exp\left(-\frac{s}{T^2}\right)$ and $\exp\left(-\frac{s}{T^2}\right)$:
\begin{eqnarray}\label{QCDN}
2M_{-}\lambda_{-}^2\exp\left( -\frac{M_{-}^2}{T^2}\right)
&=& \int_{4m_c^2}^{s_0}ds \left[\sqrt{s}\rho^1_{QCD}(s)+\rho^0_{QCD}(s)\right]\exp\left( -\frac{s}{T^2}\right)\, .
\end{eqnarray}

We differentiate   Eq.\eqref{QCDN} in regard to  $\tau=\frac{1}{T^2}$, then delete the
 pole residues $\lambda_{-}$ by adopting  a fraction  to  get  the QCD sum rules for the molecule  masses,
 \begin{eqnarray}\label{QCDSR-M}
 M^2_{-} &=& \frac{-\frac{d}{d \tau}\int_{4m_c^2}^{s_0}ds \,\left[\sqrt{s}\,\rho^1_{QCD}(s)+\,\rho^0_{QCD}(s)\right]\exp\left(- \tau s\right)}{\int_{4m_c^2}^{s_0}ds \left[\sqrt{s}\,\rho_{QCD}^1(s)+\,\rho^0_{QCD}(s)\right]\exp\left( -\tau s\right)}\, .
 \end{eqnarray}

\section{Numerical results and discussions}
At the beginning points, we take  the conventional (or commonly used) values of the  vacuum condensates
$\langle\bar{q}q \rangle=-(0.24\pm 0.01\, \rm{GeV})^3$,  $\langle\bar{s}s \rangle=(0.8\pm0.1)\langle\bar{q}q \rangle$,
 $\langle\bar{q}g_s\sigma G q \rangle=m_0^2\langle \bar{q}q \rangle$, $\langle\bar{s}g_s\sigma G s \rangle=m_0^2\langle \bar{s}s \rangle$,
$m_0^2=(0.8 \pm 0.1)\,\rm{GeV}^2$, $\langle \frac{\alpha_s
GG}{\pi}\rangle=0.012\pm0.004\,\rm{GeV}^4$    at the energy scale  $\mu=1\, \rm{GeV}$
\cite{PRT85,SVZ79,ColangeloReview}, and  take the $\overline{MS}$ masses $m_{c}(m_c)=(1.275\pm0.025)\,\rm{GeV}$
 and $m_s(\mu=2\,\rm{GeV})=(0.095\pm0.005)\,\rm{GeV}$
 from the Particle Data Group \cite{PDG}.
Then,  we take account of
the energy-scale dependence of  all the input parameters  \cite{Narison-mix},
 \begin{eqnarray}
 \langle\bar{q}q \rangle(\mu)&=&\langle\bar{q}q\rangle({\rm 1 GeV})\left[\frac{\alpha_{s}({\rm 1 GeV})}{\alpha_{s}(\mu)}\right]^{\frac{12}{33-2n_f}}\, , \nonumber\\
 \langle\bar{s}s \rangle(\mu)&=&\langle\bar{s}s \rangle({\rm 1 GeV})\left[\frac{\alpha_{s}({\rm 1 GeV})}{\alpha_{s}(\mu)}\right]^{\frac{12}{33-2n_f}}\, , \nonumber\\
 \langle\bar{q}g_s \sigma Gq \rangle(\mu)&=&\langle\bar{q}g_s \sigma Gq \rangle({\rm 1 GeV})\left[\frac{\alpha_{s}({\rm 1 GeV})}{\alpha_{s}(\mu)}\right]^{\frac{2}{33-2n_f}}\, ,\nonumber\\
  \langle\bar{s}g_s \sigma Gs \rangle(\mu)&=&\langle\bar{s}g_s \sigma Gs \rangle({\rm 1 GeV})\left[\frac{\alpha_{s}({\rm 1 GeV})}{\alpha_{s}(\mu)}\right]^{\frac{2}{33-2n_f}}\, ,\nonumber\\
m_c(\mu)&=&m_c(m_c)\left[\frac{\alpha_{s}(\mu)}{\alpha_{s}(m_c)}\right]^{\frac{12}{33-2n_f}} \, ,\nonumber\\
m_s(\mu)&=&m_s({\rm 2GeV} )\left[\frac{\alpha_{s}(\mu)}{\alpha_{s}({\rm 2GeV})}\right]^{\frac{12}{33-2n_f}}\, ,\nonumber\\
\alpha_s(\mu)&=&\frac{1}{b_0t}\left[1-\frac{b_1}{b_0^2}\frac{\log t}{t} +\frac{b_1^2(\log^2{t}-\log{t}-1)+b_0b_2}{b_0^4t^2}\right]\, ,
\end{eqnarray}
  where $t=\log \frac{\mu^2}{\Lambda^2}$, $b_0=\frac{33-2n_f}{12\pi}$, $b_1=\frac{153-19n_f}{24\pi^2}$, $b_2=\frac{2857-\frac{5033}{9}n_f+\frac{325}{27}n_f^2}{128\pi^3}$,  $\Lambda=213\,\rm{MeV}$, $296\,\rm{MeV}$  and  $339\,\rm{MeV}$ for the quark flavors  $n_f=5$, $4$ and $3$, respectively  \cite{PDG,Narison-mix}, and evolve them  from the energy scales $\mu=1\,\rm{GeV}$, $m_c$ and $2\,\rm{GeV}$ to a particular uniform energy scale $\mu$  in the QCD sum rules for a molecular state to extract the hadron mass.

In this work, we explore  the lowest hidden-charm  molecular  states without strange, with strange, and with double  strange,  it is better to choose the quark flavor numbers $n_f=4$, and evolve all the input parameters to the particular energy scales  $\mu$, which satisfy the modified  energy scale formula
$\mu=\sqrt{M^2_{X/Y/Z/P}-(2{\mathbb{M}}_c)^2}-k\,\mathbb{M}_s$ with the effective $c$-quark mass ${\mathbb{M}}_c=1.85\pm0.01\,\rm{GeV}$ and  effective $s$-quark mass $\mathbb{M}_s=0.2\,\rm{GeV}$, the subscripts $X$, $Y$, $Z$ and $P$ denote the exotic states with hidden-charm,  we take  account of  the light-flavor $SU(3)$ mass-breaking effects via counting the $s$-quark numbers $k=0$, $1$ and $2$ to assess the impact on choosing the energy scales \cite{WZG-hidden-charm-mole,Wang-tetraquark-QCDSR-2}.

In the hidden-charm (or hidden-bottom) four- and five-quark systems $Q\bar{Q}q \bar{q}^\prime$ and $Q\bar{Q}qq^{\prime}q^{\prime\prime}$, we discriminate the heavy and light degree's of freedoms explicitly, and describe them as $2{\mathbb{M}}_Q$ and $\mu+k\,\mathbb{M}_s$, respectively. We assume that the hadron masses satisfy a Regge-trajectory-like relation,
\begin{eqnarray}
M^2_{X/Y/Z/P}&=&(\mu+k\,\mathbb{M}_s)^2+C\, ,
\end{eqnarray}
where the constants $C=4{\mathbb{M}}_Q^2$,
and fit the effective masses ${\mathbb{M}}_Q$ and $\mathbb{M}_s$ by  the QCD sum rules themselves. Direct and explicit  calculations indicate that the ${\mathbb{M}}_Q$ and $\mathbb{M}_s$ have universal values and work
 well for all the exotic $X$, $Y$, $Z$ and $P$ states. We only use the universal parameters ${\mathbb{M}}_Q$ and $\mathbb{M}_s$ to determine the appropriate energy scales $\mu$ of the QCD spectral densities in a self-consistent way. While in the QCD spectral densities, we take the $\overline{MS}$ (modified minimal subtraction scheme) quark  masses.
The modified energy scale formula serves as a powerful and useful constraint to obey. On the other hand, if we set 
\begin{eqnarray}
M^2_{X/Y/Z/P}&=&(\mu+k\,\mathbb{M}_s+2\mathbb{M}_Q)^2\, ,
\end{eqnarray}
and take the best energy scales $\mu=1.3\,\rm{GeV}$ and $2.2\,\rm{GeV}$ for the $Z_c(3900)$ and $P_c(4312)$ respectively 
as the input parameters \cite{WZG-hidden-charm-mole,WZG-WXW-IsoSpin}, we obtain the effective $c$-quark mass 
${\mathbb{M}}_c=1.30\,\rm{GeV}$ and $1.06\,\rm{GeV}$, respectively, no uniform/self-consistent parameter can be reached.

We search for the suitable  Borel parameters and continuum threshold parameters to obey  the two elementary  criteria of the QCD sum rules (pole dominance and convergence of the operator product expansion play an essential role to warrant reliability) via trial and error, in fact, it is not easy to achieve such requirements  for
multiquark states. As the spectrum of the exotic states are unclear, we have no robust guides to choose the continuum thresholds, the two  criteria manifest themselves in this aspect.
Then we acquire  the Borel windows and continuum threshold parameters, therefore the optimal  energy scales of the QCD spectral densities and pole contributions of the ground states, which are all presented   plainly in Table \ref{Borel-pole}.

\begin{figure}
\centering
\includegraphics[totalheight=6cm,width=9cm]{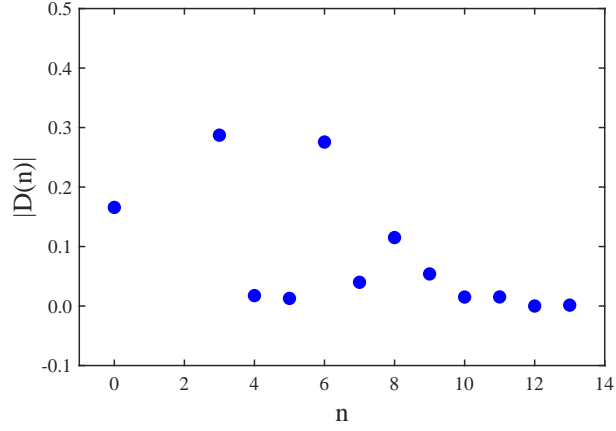}
  \caption{ The absolute values of the $D(n)$  with  central values of all the parameters for the     $\bar{D}\Xi_c$  molecular state having the isospin $(I,I_3)=(0,0)$. }\label{massDSigma-Dn}
\end{figure}

From the table, we can see clearly that the contributions from the ground states are about or slightly larger than $(40-60)\%$,  the pole dominance criterion  is satisfied very well, we choose the uniform pole contributions in all the channels to assess the reliability, if the predictions are reliable in one channel, then they are reliable in another channel, vice versa. The normalized contributions of the  condensates of dimension $n$  are defined by,
\begin{eqnarray}\label{Dn}
D(n)&=& \frac{  \int_{4m_c^2}^{s_0} ds\,\rho_{n}(s)\,\exp\left(-\frac{s}{T^2}\right)}{\int_{4m_c^2}^{s_0} ds \,\rho(s)\,\exp\left(-\frac{s}{T^2}\right)}\, ,
\end{eqnarray}
as we choose the spectral densities $\rho(s)\Theta(s-s_0)$ to approximate the continuum states, where  the $\rho_{n}(s)$ represent the terms involving the condensates of dimension $n$ in the total QCD spectral densities  $\rho(s)=\sqrt{s}\,\rho^1_{QCD}(s)+\,\rho^0_{QCD}(s)$. In calculations, we observe that the normalized contributions $D(6)$ serve  as a milestone, in all the channels, if we choose the same Borel parameter $T^2$,   the  absolute  values  $|D(n)|$ with $n\geq 6$ decrease  monotonically  and quickly with the increase of the $n$ (except that the values $|D(7)|$ are very small), and  the values $|D(13)|\ll 1\%$, the convergent behavior of the operator product expansion  is very good. In Fig.\ref{massDSigma-Dn}, we plot the absolute values of the $D(n)$  with central values of all the parameters for the  $\bar{D}\Xi_c$  molecular state having the isospin $(I,I_3)=(0,0)$ as an example. For reader's convenience, we present the lengthy QCD spectral densities in the Appendix.  

\begin{figure}
\centering
\includegraphics[totalheight=5cm,width=7cm]{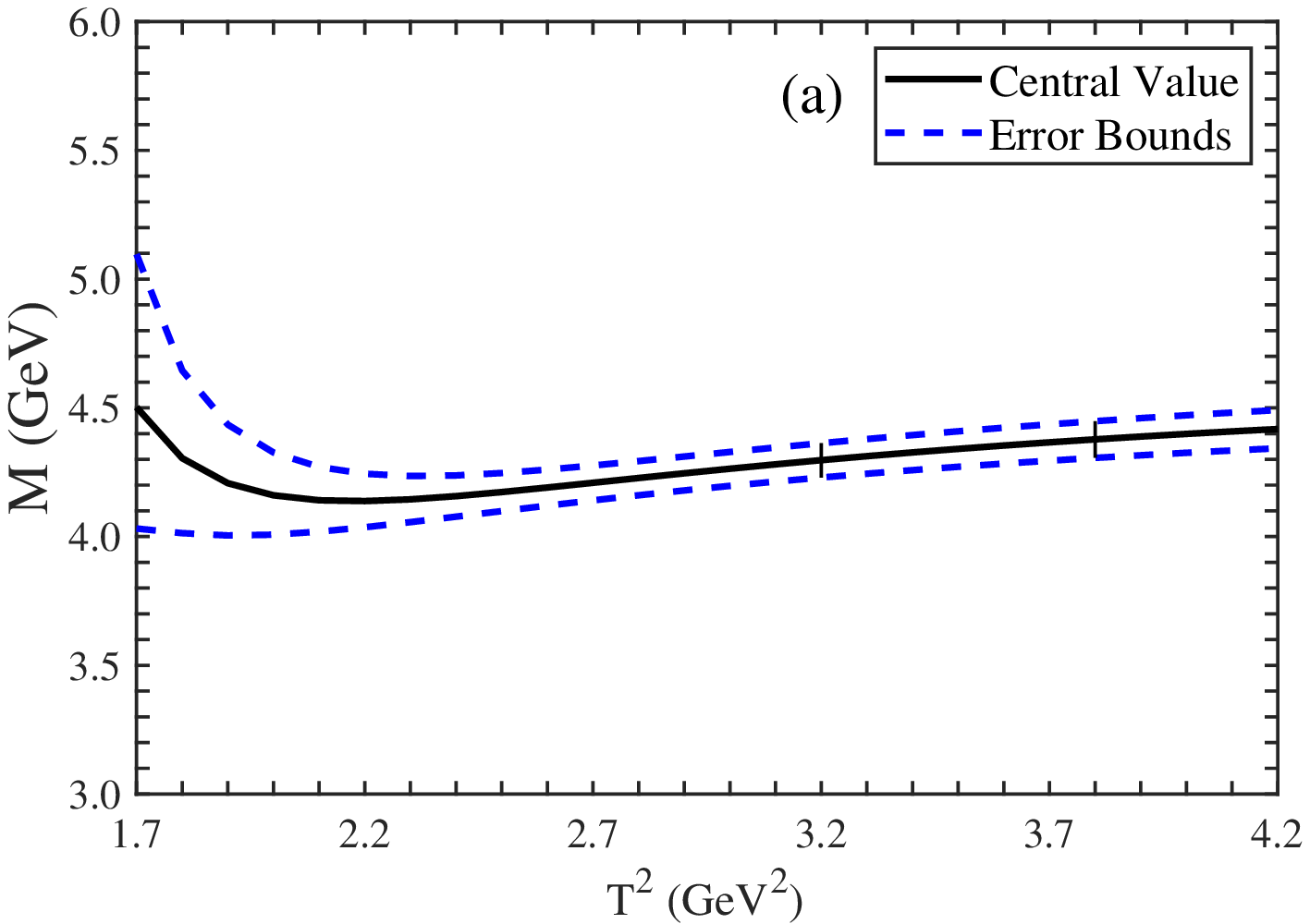}
\includegraphics[totalheight=5cm,width=7cm]{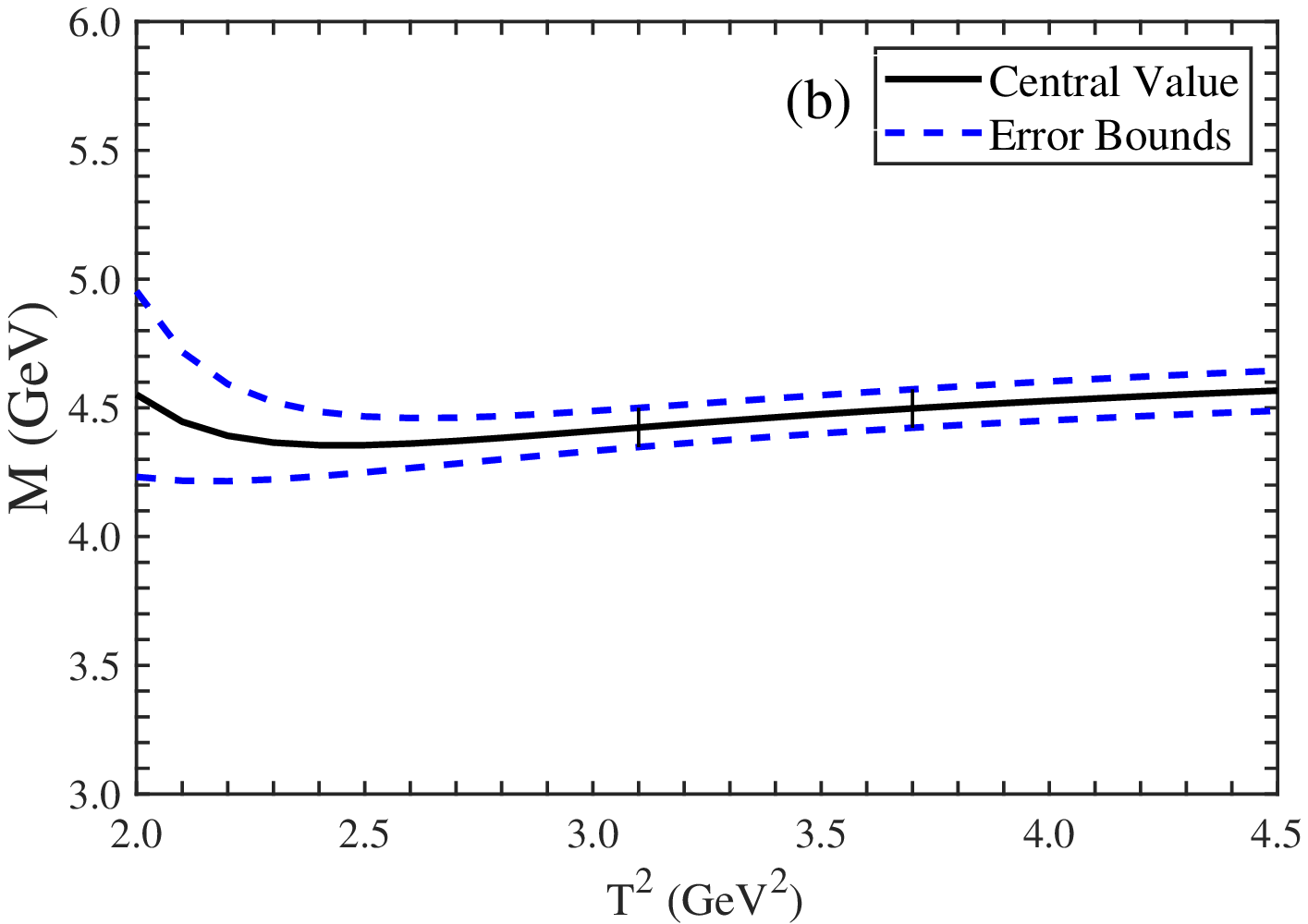}
  \caption{ The masses of the $\bar{D}\Xi_c$  molecular  states with variations of the Borel parameters $T^2$, where the (a) and (b) denote  the isospins $(0,0)$ and $(1,0)$, respectively.    }\label{massDSigma-Borel1}
\end{figure}

\begin{figure}
\centering
\includegraphics[totalheight=5cm,width=7cm]{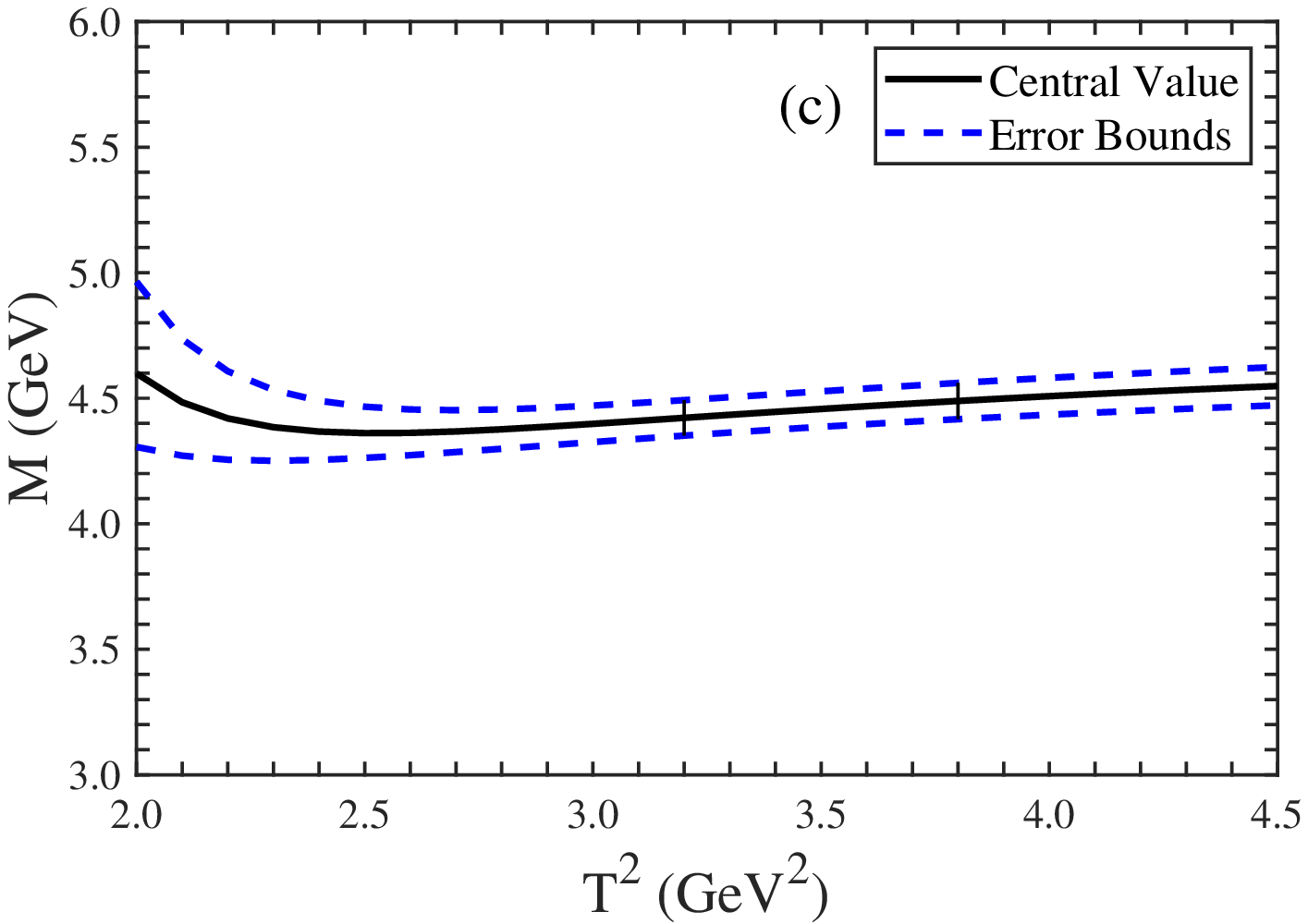}
\includegraphics[totalheight=5cm,width=7cm]{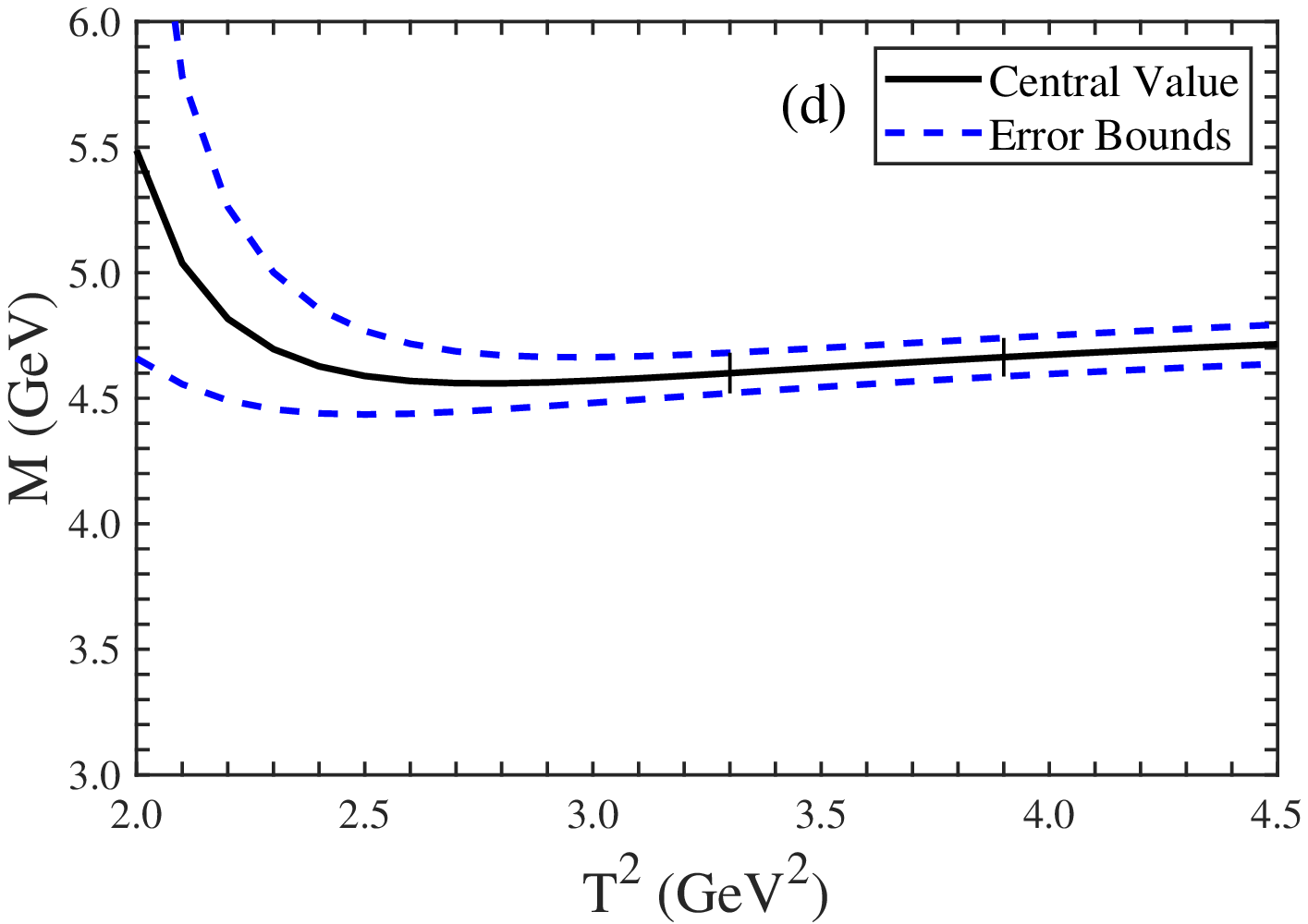}
  \caption{ The masses of the $\bar{D}^{*}\Xi_c$  molecular  states with variations of the Borel parameters $T^2$, where the (c) and (d) denote  the isospins $(0,0)$ and $(1,0)$, respectively.    }\label{massDSigma-Borel2}
\end{figure}

At the last step, we take  account of all uncertainties  of the input  parameters including the quark masses, vacuum  condensates, Borel parameters, continuum threshold parameters,
and acquire   the masses and pole residues of
 the   hidden-charm  molecular states without  strange, with strange and with double strange, and present them  explicitly in Table \ref{mass-pole-residue-tab} and Figs.\ref{massDSigma-Borel1}-\ref{massDSigma-Borel2}. From Tables \ref{Borel-pole}-\ref{mass-pole-residue-tab}, we can see clearly that the modified energy scale formula  $\mu=\sqrt{M^2_{X/Y/Z/P}-(2{\mathbb{M}}_c)^2}-k\,\mathbb{M}_s$    with the  $s$-quark numbers $k=0$, $1$ and $2$ is satisfied very well  \cite{WZG-hidden-charm-mole}.
 In Figs.\ref{massDSigma-Borel1}-\ref{massDSigma-Borel2}, we plot the masses of the $\bar{D}\Xi_c$ and $\bar{D}^*\Xi_c$ molecular states with the isospins $(I,I_3)=(0,0)$ and $(1,0)$  with variations of the Borel parameters at much larger ranges  than the Borel windows, which site  between the two short perpendicular lines. There appear very flat platforms in the Borel windows indeed, the uncertainties originate  from the Borel parameters can be ignored safely, which is congruous with the supplementary nature of the $T^2$, it is reliable to extract the molecule masses.


  The present investigations with the same constraints  indicate that there maybe exist the $\bar{D}\Xi_c$ and $\bar{D}^*\Xi_c$ molecular states  with the isospin  $(I,I_3)=(0,0)$, which lie near (irrespective slightly above or below) the corresponding charmed meson-baryon thresholds, respectively. While the $\bar{D}\Xi_c$ and $\bar{D}^*\Xi_c$ molecular states  with the isospin  $(I,I_3)=(1,0)$, the  $\bar{D}\Lambda_c$, $\bar{D}_s\Xi_c$, $\bar{D}^*\Lambda_c$ and $\bar{D}_s^*\Xi_c$ molecular states  with the isospin  $(I,I_3)=(\frac{1}{2},\frac{1}{2})$, and the $\bar{D}_s\Lambda_c$ and  $\bar{D}^*_s\Lambda_c$ molecular stats  with the isospin $(I,I_3)=(0,0)$, lie above  the corresponding charmed meson-baryon thresholds, they might be the charmed meson-baryon resonances and would have much larger widths than the $P_{cs}(4338)$ and $P_{cs}(4459)$.

 The present investigations
   favor identifying the $P_{cs}(4338)$ as the $\bar{D}\Xi_c$  molecular state with the spin-parity  $J^P={\frac{1}{2}}^-$ and isospin $I=0$, the observation of its cousin with the isospin $I=1$ in the $J/\psi\Sigma^0/\eta_c\Sigma^0$  invariant mass distributions   would
decipher the inner structure of the $P_{cs}(4338)$ and lead to more robust assignment.
The present investigations  also favor identifying the $P_{cs}(4459)$ as the $\bar{D}^*\Xi_c$  molecular state with the spin-parity  $J^P={\frac{3}{2}}^-$ and isospin $I=0$. In Ref.\cite{WZG-Pcs-mole}, we obtain the masses $M=4.45\pm0.12\,\rm{GeV}$ for the $\bar{D}\,\Xi_c^{\prime}$ molecular state with the   $J^P={\frac{1}{2}}^-$
and $M=4.51\pm0.11\,\rm{GeV}$ for the $\bar{D}\,\Xi_c^{*} $ molecular state  with the $J^P={\frac{3}{2}}^-$, which    favor
identifying the $P_{cs}(4459)$ as the $\bar{D}\Xi^\prime_c$  molecular state with the spin-parity  $J^P={\frac{1}{2}}^-$  and isospin $I=0$. At the present time, we cannot
  exclude the possibility  of identifying  the $P_{cs}(4459)$ as the $\bar{D}\Xi^*_c$  molecular state with the spin-parity $J^P={\frac{3}{2}}^-$  and isospin $I=0$ considering  uncertainty of the mass. Precise measurement of the mass and more experimental data on the quantum numbers, such as the spin and parity,  are still needed.  Furthermore, the observation of the $P_{cs}(4459)$'s cousin  with the isospin $I=1$ in the $J/\psi\Sigma^0/\eta_c\Sigma^0$  invariant mass distributions  is of crucial importance, and would
decipher  the inner structure of the $P_{cs}(4459)$, and lead to more robust assignment.

\begin{table}
\begin{center}
\begin{tabular}{|c|c|c|c|c|c|c|c|}\hline\hline
                         &$(I,I_3)$                   &$\mu(\rm GeV)$    &$T^2 (\rm{GeV}^2)$  &$\sqrt{s_0}(\rm{GeV})$   &pole     \\ \hline

$\bar{D}\,\Xi_c$         &$(0,0)$                     &2.1               &$3.2-3.8$           &$5.00\pm0.10$            &$(41-60)\%$  \\   \hline

$\bar{D}\,\Xi_c$         &$(1,0)$                     &2.3               &$3.1-3.7$           &$5.09\pm0.10$            &$(42-61)\%$  \\   \hline

$\bar{D}\,\Lambda_c$     &$(\frac{1}{2},\frac{1}{2})$ &2.5               &$3.2-3.8$           &$5.11\pm0.10$            &$(42-60)\%$  \\   \hline

$\bar{D}_s\,\Xi_c$       &$(\frac{1}{2},\frac{1}{2})$ &2.2               &$3.2-3.8$           &$5.15\pm0.10$            &$(41-59)\%$  \\   \hline

$\bar{D}_s\,\Lambda_c$   &$(0,0)$                     &2.3               &$3.2-3.8$           &$5.13\pm0.10$            &$(43-61)\%$  \\   \hline

$\bar{D}^*\,\Xi_c$       &$(0,0)$                     &2.3               &$3.2-3.8$           &$5.10\pm0.10$            &$(43-61)\%$  \\   \hline

$\bar{D}^*\,\Xi_c$       &$(1,0)$                     &2.6               &$3.3-3.9$           &$5.27\pm0.10$            &$(43-61)\%$  \\   \hline

$\bar{D}^*\,\Lambda_c$   &$(\frac{1}{2},\frac{1}{2})$ &2.7               &$3.3-3.9$           &$5.23\pm0.10$            &$(41-61)\%$  \\   \hline

$\bar{D}^*_s\,\Xi_c$     &$(\frac{1}{2},\frac{1}{2})$ &2.4               &$3.3-3.9$           &$5.28\pm0.10$            &$(42-59)\%$  \\   \hline

$\bar{D}^*_s\,\Lambda_c$ &$(0,0)$                     &2.4               &$3.2-3.8$           &$5.14\pm0.10$            &$(42-60)\%$  \\   \hline

\hline
\end{tabular}
\end{center}
\caption{ The  best  energy scales $\mu$, Borel windows $T^2$, continuum threshold parameters $s_0$ and
 pole contributions (pole)    for the hidden-charm   pentaquark molecular states.} \label{Borel-pole}
\end{table}

\begin{table}
\begin{center}
\begin{tabular}{|c|c|c|c|c|c|c|c|}\hline\hline
                         &$(I,I_3)$                      &$M (\rm{GeV})$     &$\lambda (10^{-3}\rm{GeV}^6)$ &Thresholds (MeV) & Assignments        \\  \hline

$\bar{D}\,\Xi_c$         &$(0,0)$                        &$4.34_{-0.07}^{+0.07}$  &$1.43^{+0.19}_{-0.18}$   &4337              & ? $P_{cs}(4338)$     \\      \hline

$\bar{D}\,\Xi_c$         &$(1,0)$                        &$4.46_{-0.07}^{+0.07}$  &$1.37^{+0.19}_{-0.18}$   &4337              &    \\      \hline

$\bar{D}\,\Lambda_c$     &$(\frac{1}{2},\frac{1}{2})$    &$4.46_{-0.08}^{+0.07}$  &$1.47^{+0.20}_{-0.18}$   &4151              &     \\  \hline

$\bar{D}_s\,\Xi_c$       &$(\frac{1}{2},\frac{1}{2})$    &$4.54_{-0.07}^{+0.07}$  &$1.58^{+0.21}_{-0.20}$   &4437              &       \\  \hline

$\bar{D}_s\,\Lambda_c$   &$(0,0)$                        &$4.48_{-0.07}^{+0.07}$  &$1.57^{+0.21}_{-0.20}$   &4255              &      \\ \hline
$\bar{D}^*\,\Xi_c$       &$(0,0)$                        &$4.46_{-0.07}^{+0.07}$  &$1.55^{+0.20}_{-0.19}$   &4479              & ? $P_{cs}(4459)$     \\      \hline

$\bar{D}^*\,\Xi_c$       &$(1,0)$                        &$4.63_{-0.08}^{+0.08}$  &$1.69^{+0.22}_{-0.21}$   &4479              &    \\      \hline

$\bar{D}^*\,\Lambda_c$   &$(\frac{1}{2},\frac{1}{2})$    &$4.59_{-0.08}^{+0.08}$  &$1.67^{+0.22}_{-0.21}$   &4293              &     \\  \hline

$\bar{D}^*_s\,\Xi_c$     &$(\frac{1}{2},\frac{1}{2})$    &$4.65_{-0.08}^{+0.08}$  &$1.66^{+0.22}_{-0.21}$   &4580              &       \\  \hline

$\bar{D}^*_s\,\Lambda_c$ &$(0,0)$                        &$4.50_{-0.07}^{+0.07}$  &$1.52^{+0.21}_{-0.19}$   &4398              &      \\ \hline

\hline
\end{tabular}
\end{center}
\caption{ The  masses and pole residues of the  pentaquark molecular states with the possible assignments.} \label{mass-pole-residue-tab}
\end{table}

\section{Conclusion}
In the present work, we extend our previous works on the pentaquark (molecular) states and  distinguish the isospins of the interpolating currents to investigate the $\bar{D}\,\Xi_c$, $\bar{D}\,\Lambda_c$, $\bar{D}_s\,\Xi_c$ $\bar{D}_s\,\Lambda_c$, $\bar{D}^*\,\Xi_c$, $\bar{D}^*\,\Lambda_c$, $\bar{D}^*_s\,\Xi_c$ and $\bar{D}^*_s\,\Lambda_c$  molecular states  without strange, with strange  and  with double strange  in the framework of  the QCD sum rules in details. Here the $D$, $D_s$, $D^*$ and $D_s^*$ stand for the color-singlet clusters having the same quantum numbers as the physical  mesons, the $\Lambda_c$ and $\Xi_c$ stand for the color-singlet clusters having the same quantum numbers as the physical  ground state flavor-antitriplet charmed baryons. As the charmed baryons in flavor antitriplet have smaller masses than that in flavor sextet having the same valence quarks, we expect to acquire the lowest molecular states (to be more precise, the color-singlet-color-singlet type pentaquark states).

We accomplish  the operator product expansion   up to   the vacuum condensates of dimension $13$  consistently, and choose the best energy scales of the QCD spectral densities with the help of the modified energy scale formula, which plays a crucial important role in matching  the two fundamental criteria of the QCD sum rules,
and we acquire  the masses and pole residues of those  molecular states.
 The present investigations   favor identifying the $P_{cs}(4338)$ ($P_{cs}(4459)$) as the $\bar{D}\Xi_c$ ($\bar{D}^*\Xi_c$) molecular state  with the spin-parity   $J^P={\frac{1}{2}}^-$ (${\frac{3}{2}}^-$) and the isospin $(I,I_3)=(0,0)$, which are in congruous with the decays to the final states  $J/\psi\Lambda$,
 the observation of their cousins with the isospin $(I,I_3)=(1,0)$ in the $J/\psi\Sigma^0/\eta_c\Sigma^0$ invariant mass distributions would
decipher the inner structures of the $P_{cs}(4338)$ and $P_{cs}(4459)$, and lead to more robust assignments.
While in the picture of diquark-diquark-antiquark type pentaquark states, the $P_{cs}(4338)$ cannot find its position, the $P_{cs}(4459)$ can be identified as the strange partner of the $P_{c}(4312)$ tentatively.

More experimental data are still needed to reach final assignments. Furthermore,  we make predictions for other pentaquark molecular states with the isospins $(I,I_3)=(\frac{1}{2},\frac{1}{2})$ and $(0,0)$, which lie above the corresponding charmed meson-baryon thresholds, and it is better to call them resonances, and they would have much larger widths than the $P_{cs}(4338)$ and $P_{cs}(4459)$. All the predictions can be confronted to the experimental data in the future.

\section*{Appendix}

The detailed QCD spectral densities for the current $J_{(0,0)}^{\bar{D}\Xi_c}(x)$,

\begin{eqnarray}
\notag
\rho_{QCD}^1(s)&=&\sum\limits_{n}\left[\rho^1_{a}(n)+\rho^1_{b}(n)+\rho^1_{c}(n)\delta(s-\widetilde{m}_c^2)+\rho^1_{d}(n)
\delta(s-\overline{m}_c^2)\right]\, ,\\
\notag \rho_{QCD}^0(s)&=&\sum\limits_{n}\left[\rho^0_{a}(n)+\rho^0_{b}(n)+\rho^0_{c}(n)\delta(s-\widetilde{m}_c^2)+
\rho^0_{d}(n)\delta(s-\overline{m}_c^2)\right]\, ,
\end{eqnarray}
 where the $a$, $b$, $c$ and $d$ refer to four types of integrals, $n$ are the dimension of the condensates.
In those integrals, we introduce the notations  $\widetilde{m}_c^2=\frac{m_c^2}{y(1-y)}$, $\overline{m}_c^2=\frac{(y+z)m_c^2}{y z}$, $\xi=y+z-1$, $\zeta=1-y$ and $\omega=s-\overline{m}_c^2$.
 For the types $a$ and $b$, $y_i=\frac{1}{2}\left(1-\sqrt{1-4m_c^2/s}\right)$, $y_f=\frac{1}{2}\left(1+\sqrt{1-4m_c^2/s}\right)$ and $z_i=\frac{y m_c^2}{y s-m_c^2}$. For the types $c$ and $d$, $y_i=0$, $y_f=1$ and $z_i=0$.

The $a$ type integrals:
\begin{eqnarray}
\notag \rho^1_a(8)&=&\frac{m_s m_c \left[ 9 \langle\bar{q}g_s\sigma Gq\rangle \langle\bar{s}s\rangle-39 \langle\bar{q}q\rangle \langle\bar{q}g_s\sigma Gq\rangle + 7 \langle\bar{q}q\rangle \langle\bar{s}g_s\sigma Gs\rangle\right]}{9216\pi^4}  \int_{y_i}^{y_f}dy \, \zeta \, ,
\notag
\end{eqnarray}

\begin{eqnarray}
\notag \rho^1_a(9)&=& -\frac{13 m_c \langle\bar{s}s\rangle \langle\bar{q}q\rangle^2}{1152\pi^2} \int_{y_i}^{y_f}dy  \, \zeta-\frac{m_c g_s^2 \langle\bar{q}q\rangle \left[7 \langle\bar{q}q\rangle^2 + \langle\bar{q}q\rangle \langle\bar{s}s\rangle + 7 \langle\bar{s}s\rangle^2\right]}{62208\pi^4}  \int_{y_i}^{y_f}dy \, \zeta  \\
\notag &&+\frac{ m_s \langle\bar{s}s\rangle \langle\bar{q}q\rangle^2}{768\pi^2}  \int_{y_i}^{y_f}dy \, \zeta y
-\frac{m_s g_s^2 \langle\bar{q}q\rangle^2 \left[14 \langle\bar{q}q\rangle - 13 \langle\bar{s}s\rangle\right]}{41472\pi^4}  \int_{y_i}^{y_f}dy \, \zeta y  \, ,
\end{eqnarray}

\begin{eqnarray}
\notag \rho^1_a(10)&=&  \frac{11 \langle\bar{q}g_s\sigma Gq\rangle \langle\bar{s}g_s\sigma Gs\rangle}{8192\pi^4}\int_{y_i}^{y_f}dy \, \zeta y
+\frac{\langle g_s^2GG\rangle \langle\bar{q}q\rangle \left[\langle\bar{q}q\rangle + 14 \langle\bar{s}s\rangle\right]}{36864\pi^4}  \int_{y_i}^{y_f}dy  \, \zeta y  \, ,
\end{eqnarray}

\begin{eqnarray}
\notag \rho^0_a(8)&=&- \frac{m_s m_c^2 \left[39 \langle\bar{q}q\rangle \langle\bar{q}g_s\sigma Gq\rangle - 9 \langle\bar{q}g_s\sigma Gq\rangle \langle\bar{s}s\rangle - 7 \langle\bar{q}q\rangle \langle\bar{s}g_s\sigma Gs\rangle\right]}{9216\pi^4}   \int_{y_i}^{y_f}dy   \, ,
\end{eqnarray}

\begin{eqnarray}
\notag \rho^0_a(9) &=&-  \frac{13 m_c^2\langle\bar{q}q\rangle^2 \langle\bar{s}s\rangle}{1152\pi^2} \int_{y_i}^{y_f}dy   - \frac{m_c^2  g_s^2 \langle\bar{q}q\rangle \left[7 \langle\bar{q}q\rangle^2 + \langle\bar{q}q\rangle \langle\bar{s}s\rangle + 7 \langle\bar{s}s\rangle^2\right]}{62208\pi^4}  \int_{y_i}^{y_f}dy   \\
\notag &&+  \frac{ m_s m_c \langle\bar{q}q\rangle^2 \langle\bar{s}s\rangle}{1152\pi^2} \int_{y_i}^{y_f}dy \, y    + \frac{m_s m_c  g_s^2 \langle\bar{q}q\rangle^2 \left[-14 \langle\bar{q}q\rangle + 13 \langle\bar{s}s\rangle\right]}{62208\pi^4}   \int_{y_i}^{y_f}dy  \,y \, ,
\end{eqnarray}

\begin{eqnarray}
\notag \rho^0_a(10)&=&  \frac{11 m_c \langle\bar{q}g_s\sigma Gq\rangle \langle\bar{s}g_s\sigma Gs\rangle}{12288\pi^4} \int_{y_i}^{y_f}dy  \,y  + \frac{m_c\langle g_s^2GG\rangle \langle\bar{q}q\rangle \left[\langle\bar{q}q\rangle + 14 \langle\bar{s}s\rangle\right]}{55296\pi^4}  \int_{y_i}^{y_f}dy \, y \, .
\end{eqnarray}

The $b$ type integrals:
\begin{eqnarray}
\notag \rho_b^1(0)&=& \frac{13}{1572864 \pi^8}  \int_{y_i}^{y_f}dy\int_{z_i}^{\zeta}dz  \,y z \xi^4 \left(\frac{3\omega^5}{5}+s \omega^4\right) -  \frac{ m_s m_c}{393216\pi^8}  \int_{y_i}^{y_f}dy\int_{z_i}^{\zeta}dz \,  z \xi^3\omega^4 \, ,
\notag
\end{eqnarray}

\begin{eqnarray}
\notag \rho_b^1(3)&=& -\frac{m_c \left[14 \langle\bar{q}q\rangle + \langle\bar{s}s\rangle\right]}{24576 \pi^6 }
 \int_{y_i}^{y_f}dy\int_{z_i}^{\zeta}dz  \,z \xi^2\omega^3  + \frac{m_s\left[13 \langle\bar{s}s\rangle-28 \langle\bar{q}q\rangle\right]}{16384 \pi^6} \int_{y_i}^{y_f}dy\int_{z_i}^{\zeta}dz  \,y z \xi^2\left(s+\omega\right) \omega^2 \, ,
\end{eqnarray}

\begin{eqnarray}
\notag \rho_b^1(4)&=& -\frac{13 m_c^2\langle g_s^2GG\rangle}{786432 \pi^8} \int_{y_i}^{y_f}dy\int_{z_i}^{\zeta}dz  \,\frac{z \xi^4}{y^2} \left(\omega^2+\frac{2 s \omega}{3}\right)- \frac{29 \langle g_s^2GG\rangle}{1572864 \pi^8 }  \int_{y_i}^{y_f}dy\int_{z_i}^{\zeta}dz \, z \xi^3 \left(\omega^3+s \omega^2\right)\\
\notag &&+  \frac{\langle g_s^2GG\rangle}{32768\pi^8} \int_{y_i}^{y_f}dy\int_{z_i}^{\zeta}dz  \, y z \xi^2 \left(\omega^3+s \omega^2\right) + \frac{m_s m_c^3 \langle g_s^2GG\rangle}{1179648 \pi^8 }
  \int_{y_i}^{y_f}dy\int_{z_i}^{\zeta}dz  \, \frac{\xi^3\omega}{y^2} \\
\notag &&+\frac{m_s m_c \langle g_s^2GG\rangle}{786432\pi^8} \int_{y_i}^{y_f}dy\int_{z_i}^{\zeta}dz  \, \frac{z \xi^3}{y^2} \left(y \omega^2+\frac{2 s y \omega}{3}-\frac{2 \omega^2}{3}\right)\\
\notag &&+ \frac{m_s m_c \langle g_s^2GG\rangle}{131072\pi^8} \int_{y_i}^{y_f}dy\int_{z_i}^{\zeta}dz  \left(\frac{z \xi^2\omega^2}{y}-z \xi\omega^2\right) \, ,
\end{eqnarray}

\begin{eqnarray}
\notag \rho_b^1(5)&=& -\frac{m_c \langle\bar{q}g_s\sigma Gq\rangle}{16384\pi^6}  \int_{y_i}^{y_f}dy\int_{z_i}^{\zeta}dz  \left(11 z \xi\omega^2+\frac{14 z \xi^2\omega^2}{y}\right) \\
\notag &&+ \frac{ m_s}{24576\pi^6} \left[-36 \langle\bar{q}g_s\sigma Gq\rangle + 13 \langle\bar{s}g_s\sigma Gs\rangle\right] \int_{y_i}^{y_f}dy\int_{z_i}^{\zeta}dz  \, y z \xi \left(\frac{3 \omega^2}{2}+s \omega\right)\\
\notag &&- \frac{m_s \langle\bar{q}g_s\sigma Gq\rangle}{8192\pi^6} \int_{y_i}^{y_f}dy\int_{z_i}^{\zeta}dz  \, z \xi^2 \left(\frac{3 \omega^2}{2}+s \omega\right)\, ,
\end{eqnarray}

\begin{eqnarray}
\notag \rho_b^1(6)&=& -\frac{\langle\bar{q}q\rangle\left[\langle\bar{q}q\rangle+14\langle\bar{s}s\rangle\right]}{1024\pi^4}  \int_{y_i}^{y_f}dy\int_{z_i}^{\zeta}dz  \, y z \xi  \left(\frac{2 s \omega}{3 }+ \omega^2\right)\\
\notag &&-\frac{13\left[2 g_s^2 \langle\bar{q}q\rangle^2+ g_s^2 \langle\bar{s}s\rangle^2\right]}{110592\pi^6}  \int_{y_i}^{y_f}dy\int_{z_i}^{\zeta}dz  \, y z \xi  \left(\frac{2 s \omega}{3 }+ \omega^2\right)\\
\notag &&+ \frac{ m_s m_c \langle\bar{q}q\rangle \left[13 \langle\bar{q}q\rangle - 7 \langle\bar{s}s\rangle\right]}{1536\pi^4}   \int_{y_i}^{y_f}dy\int_{z_i}^{\zeta}dz \,z\omega + \frac{ m_s m_c g_s^2 \langle\bar{q}q\rangle^2}{82944\pi^6}   \int_{y_i}^{y_f}dy\int_{z_i}^{\zeta}dz \,z\omega \, ,
\end{eqnarray}

\begin{eqnarray}
\notag \rho_b^1(7)&=& \frac{m_c^3 \langle g_s^2GG\rangle \left[14 \langle\bar{q}q\rangle + \langle\bar{s}s\rangle\right]}{294912 \pi^6 } \int_{y_i}^{y_f}dy\int_{z_i}^{\zeta}dz  \, \frac{\xi^2}{y^2}\\
\notag &&+ \frac{m_c\langle g_s^2GG\rangle \left[14 \langle\bar{q}q\rangle + \langle\bar{s}s\rangle\right]}{98304\pi^6}  \int_{y_i}^{y_f}dy\int_{z_i}^{\zeta}dz  \, \frac{z \xi^2}{y^2} \left(y \omega-\frac{2\omega}{3}+\frac{s y}{3}\right)\\
\notag &&+ \frac{m_c\langle g_s^2GG\rangle \left[3 \langle\bar{q}q\rangle + 2 \langle\bar{s}s\rangle\right]}{49152\pi^6}  \int_{y_i}^{y_f}dy\int_{z_i}^{\zeta}dz  \, \frac{z \xi\omega}{y}\\
\notag && -\frac{m_c\langle g_s^2GG\rangle \left[56 \langle\bar{q}q\rangle + 7\langle\bar{s}s\rangle\right]}{294912\pi^6}  \int_{y_i}^{y_f}dy\int_{z_i}^{\zeta}dz  \, z\omega\\
\notag &&- \frac{m_s m_c^2 \langle g_s^2GG\rangle \left[-28 \langle\bar{q}q\rangle + 13 \langle\bar{s}s\rangle\right]}{98304\pi^6}  \int_{y_i}^{y_f}dy\int_{z_i}^{\zeta}dz  \, \frac{z \xi^2}{y^2} \\
\notag && +\frac{7 m_s\langle g_s^2GG\rangle \left[2 \langle\bar{q}q\rangle - \langle\bar{s}s\rangle\right]}{49152\pi^6}  \int_{y_i}^{y_f}dy\int_{z_i}^{\zeta}dz  \, z \xi \left(s+3 \omega\right)\\
\notag &&+  \frac{ m_s \langle g_s^2GG\rangle \left[3\langle\bar{s}s\rangle-14\langle\bar{q}q\rangle\right]}{294912\pi^6}
 \int_{y_i}^{y_f}dy\int_{z_i}^{\zeta}dz  \, y z \left(s +3 \omega\right)\, ,
\end{eqnarray}

\begin{eqnarray}
\notag \rho_b^1(8)&=& -\frac{12 \langle\bar{q}g_s\sigma Gq\rangle \langle\bar{s}s\rangle + \langle\bar{q}q\rangle \left[\langle\bar{q}g_s\sigma Gq\rangle + 13 \langle\bar{s}g_s\sigma Gs\rangle\right]}{6144\pi^4} \int_{y_i}^{y_f}dy\int_{z_i}^{\zeta}dz  \, y z \left(s+3 \omega\right)\\
\notag && -\frac{2 \langle\bar{q}g_s\sigma Gq\rangle \langle\bar{s}s\rangle + \langle\bar{q}q\rangle \left[2 \langle\bar{q}g_s\sigma Gq\rangle + \langle\bar{s}g_s\sigma Gs\rangle\right]}{6144\pi^4} \int_{y_i}^{y_f}dy\int_{z_i}^{\zeta}dz  \, z \xi \left(s+3 \omega\right)\\
\notag &&+ \frac{7 m_s m_c \langle\bar{q}g_s\sigma Gq\rangle \left[2 \langle\bar{q}q\rangle - \langle\bar{s}s\rangle\right]}{3072\pi^4}  \int_{y_i}^{y_f}dy\int_{z_i}^{\zeta}dz  \, \frac{z}{y}\, ,
\end{eqnarray}

\begin{eqnarray}
\notag \rho_b^1(10)&=& \frac{\langle g_s^2GG\rangle \langle\bar{q}q\rangle \left[\langle\bar{q}q\rangle + 20 \langle\bar{s}s\rangle\right]}{24576\pi^4}  \int_{y_i}^{y_f}dy\int_{z_i}^{\zeta}dz  \,z \\
\notag &&- \frac{\langle\bar{q}g_s\sigma Gq\rangle \left[24 \langle\bar{q}g_s\sigma Gq\rangle + 35 \langle\bar{s}g_s\sigma Gs\rangle\right]}{196608\pi^4}  \int_{y_i}^{y_f}dy\int_{z_i}^{\zeta}dz  \, z \, ,
\end{eqnarray}

\begin{eqnarray}
\notag \rho_b^0(0)&=&   \frac{13 m_c}{1572864 \pi^8}  \int_{y_i}^{y_f}dy\int_{z_i}^{\zeta}dz  \, y \xi^4 \left(\frac{2 \omega^5}{5}+s \omega^4\right)-  \frac{ m_s m_c^2}{393216\pi^8}  \int_{y_i}^{y_f}dy\int_{z_i}^{\zeta}dz  \, \xi^3\omega^4  \, ,
\end{eqnarray}

\begin{eqnarray}
\notag \rho_b^0(3)&=&  -\frac{m_c^2 \left[14 \langle\bar{q}q\rangle + \langle\bar{s}s\rangle\right]}{24576\pi^6}
 \int_{y_i}^{y_f}dy\int_{z_i}^{\zeta}dz  \, \xi^2\omega^3 \\
\notag &&- \frac{m_c m_s \left[28 \langle\bar{q}q\rangle - 13 \langle\bar{s}s\rangle\right]}{16384\pi^6} \int_{y_i}^{y_f}dy\int_{z_i}^{\zeta}dz  \, y \xi^2 \left(\frac{2\omega^3}{3}+s \omega^2\right)\, ,
\end{eqnarray}

\begin{eqnarray}
\notag \rho_b^0(4)&=&  -\frac{13 m_c^3 \langle g_s^2GG\rangle}{4718592\pi^8} \int_{y_i}^{y_f}dy\int_{z_i}^{\zeta}dz  \, \frac{\xi^4}{y^2} \left(\omega^2+s \omega\right)\\
\notag && -\frac{13 m_c \langle g_s^2GG\rangle}{4718592\pi^8} \int_{y_i}^{y_f}dy\int_{z_i}^{\zeta}dz  \, \frac{z \xi^4}{y^2} \left(s^2 y \omega+y \omega^3-s \omega^2\right)\\
\notag && -\frac{13 m_c \langle g_s^2GG\rangle}{1572864\pi^8} \int_{y_i}^{y_f}dy\int_{z_i}^{\zeta}dz  \, \frac{z \xi^4}{y^2} \left(s y \omega^2-\frac{2\omega^3}{9}\right)\\
\notag &&-  \frac{29 m_c \langle g_s^2GG\rangle}{1572864\pi^8} \int_{y_i}^{y_f}dy\int_{z_i}^{\zeta}dz  \, \xi^3 \left(\frac{2\omega^3}{3}+s \omega^2\right)\\
\notag &&+  \frac{ m_c \langle g_s^2GG\rangle}{32768\pi^8} \int_{y_i}^{y_f}dy\int_{z_i}^{\zeta}dz  \, y \xi^2 \left(\frac{2\omega^3}{3}+s \omega^2\right)\\
\notag && -\frac{m_s m_c^2 \langle g_s^2GG\rangle}{589824\pi^8} \int_{y_i}^{y_f}dy\int_{z_i}^{\zeta}dz  \, \frac{\xi^3}{y^2} \left(\zeta \omega^2-s y \omega\right)\\
\notag &&+  \frac{m_s m_c^2 \langle g_s^2GG\rangle}{131072\pi^8} \int_{y_i}^{y_f}dy\int_{z_i}^{\zeta}dz  \left(\frac{\xi^2\omega^2}{y}-\xi\omega^2\right) \, ,
\end{eqnarray}

\begin{eqnarray}
\notag \rho_b^0(5)&=&  -\frac{m_c^2 \left[14 \langle\bar{q}g_s\sigma Gq\rangle + \langle\bar{s}g_s\sigma Gs\rangle\right]}{16384\pi^6} \int_{y_i}^{y_f}dy\int_{z_i}^{\zeta}dz  \, \xi\omega^2 \\
\notag &&- \frac{7 m_c^2 \langle\bar{q}g_s\sigma Gq\rangle}{8192\pi^6} \int_{y_i}^{y_f}dy\int_{z_i}^{\zeta}dz  \, \frac{\xi^2\omega^2}{y} \\
\notag &&+ \frac{m_c^2 \left[3 \langle\bar{q}g_s\sigma Gq\rangle + \langle\bar{s}g_s\sigma Gs\rangle\right]}{16384\pi^6}  \int_{y_i}^{y_f}dy\int_{z_i}^{\zeta}dz  \, \xi\omega^2\\
\notag &&- \frac{m_s m_c \left[42 \langle\bar{q}g_s\sigma Gq\rangle - 13 \langle\bar{s}g_s\sigma Gs\rangle\right]}{24576\pi^6}  \int_{y_i}^{y_f}dy\int_{z_i}^{\zeta}dz  \, y \xi \left(\omega^2+s \omega\right)\\
\notag && -\frac{m_s m_c\langle\bar{q}g_s\sigma Gq\rangle}{8192\pi^6} \int_{y_i}^{y_f}dy\int_{z_i}^{\zeta}dz  \, \xi^2 \left(\omega^2+s \omega\right)\\
\notag && +\frac{m_s m_c \langle\bar{q}g_s\sigma Gq\rangle}{4096\pi^6} \int_{y_i}^{y_f}dy\int_{z_i}^{\zeta}dz  \, y \xi \left(\omega^2+s \omega\right)\, ,
\end{eqnarray}

\begin{eqnarray}
\notag \rho_b^0(6)&=&  -\frac{m_c \langle\bar{q}q\rangle \left[\langle\bar{q}q\rangle + 14 \langle\bar{s}s\rangle\right]}{1536\pi^4}  \int_{y_i}^{y_f}dy\int_{z_i}^{\zeta}dz  \, y \xi \left(\omega^2+s \omega\right)\\
\notag &&+ \frac{m_s m_c^2 \langle\bar{q}q\rangle \left[13 \langle\bar{q}q\rangle - 7 \langle\bar{s}s\rangle\right]}{1536\pi^4}  \int_{y_i}^{y_f}dy\int_{z_i}^{\zeta}dz  \, \omega+ \frac{m_s m_c^2 g_s^2 \langle\bar{q}q\rangle^2}{82944\pi^6}  \int_{y_i}^{y_f}dy\int_{z_i}^{\zeta}dz  \, \omega\\
\notag && -\frac{13 m_c \left[2  g_s^2 \langle\bar{q}q\rangle^2 +  g_s^2 \langle\bar{s}s\rangle^2\right]}{165888\pi^6} \int_{y_i}^{y_f}dy\int_{z_i}^{\zeta}dz  \, y \xi \left(\omega^2+s \omega\right)\, ,
\end{eqnarray}

\begin{eqnarray}
\notag \rho_b^0(7)&=&  \frac{m_c^2\langle g_s^2GG\rangle \left[14 \langle\bar{q}q\rangle + \langle\bar{s}s\rangle\right]}{73728\pi^6}  \int_{y_i}^{y_f}dy\int_{z_i}^{\zeta}dz  \, \frac{\xi^2}{y^2} \left(-\zeta\omega+\frac{s y}{2}\right)\\
\notag &&+ \frac{m_c^2\langle g_s^2GG\rangle \left[3 \langle\bar{q}q\rangle + 2 \langle\bar{s}s\rangle\right]}{49152\pi^6}  \int_{y_i}^{y_f}dy\int_{z_i}^{\zeta}dz  \, \frac{\xi\omega}{y}\\
\notag && -\frac{7m_c^2\langle g_s^2GG\rangle \left[8 \langle\bar{q}q\rangle + \langle\bar{s}s\rangle\right]}{294912\pi^6} \int_{y_i}^{y_f}dy\int_{z_i}^{\zeta}dz  \, \omega\\
\notag &&+ \frac{ m_s m_c^3\langle g_s^2GG\rangle \left[28 \langle\bar{q}q\rangle - 13 \langle\bar{s}s\rangle\right]}{294912\pi^6}  \int_{y_i}^{y_f}dy\int_{z_i}^{\zeta}dz  \left(\frac{\xi^2}{y^2}+\frac{z \xi^2}{y^3}\right) \\
\notag &&- \frac{m_s m_c\langle g_s^2GG\rangle \left[28 \langle\bar{q}q\rangle - 13 \langle\bar{s}s\rangle\right]}{196608\pi^6} \int_{y_i}^{y_f}dy\int_{z_i}^{\zeta}dz  \, \frac{z \xi^2}{y^2} \left(s +2\omega\right)\\
\notag &&+ \frac{7 m_s m_c\langle g_s^2GG\rangle \left[2 \langle\bar{q}q\rangle - \langle\bar{s}s\rangle\right]}{49152\pi^6} \int_{y_i}^{y_f}dy\int_{z_i}^{\zeta}dz  \, \xi \left(s+2\omega\right)\\
\notag &&+\frac{ m_s m_c \langle g_s^2GG\rangle \langle\bar{s}s\rangle}{98304\pi^6} \int_{y_i}^{y_f}dy\int_{z_i}^{\zeta}dz  \, y \left(s+2\omega\right)\\
\notag &&-  \frac{7 m_s m_c \langle g_s^2GG\rangle \langle\bar{q}q\rangle}{147456\pi^6} \int_{y_i}^{y_f}dy\int_{z_i}^{\zeta}dz  \, y \left(s+2\omega\right)\, ,
\end{eqnarray}

\begin{eqnarray}
\notag \rho_b^0(8)&=&  -\frac{ m_c \left\{7 \langle\bar{q}g_s\sigma Gq\rangle \langle\bar{s}s\rangle + \langle\bar{q}q\rangle \left[\langle\bar{q}g_s\sigma Gq\rangle + 7 \langle\bar{s}g_s\sigma Gs\rangle\right]\right\}}{3072\pi^4}  \int_{y_i}^{y_f}dy\int_{z_i}^{\zeta}dz \left(s y+2y\omega\right)\\
\notag && -\frac{m_c\left\{2 \langle\bar{q}g_s\sigma Gq\rangle \langle\bar{s}s\rangle + \langle\bar{q}q\rangle \left[2 \langle\bar{q}g_s\sigma Gq\rangle + \langle\bar{s}g_s\sigma Gs\rangle\right]\right\}}{6144\pi^4} \int_{y_i}^{y_f}dy\int_{z_i}^{\zeta}dz \left(\xi s+2\xi\omega\right)\\
\notag && +\frac{m_c\left\{2 \langle\bar{q}g_s\sigma Gq\rangle \langle\bar{s}s\rangle + \langle\bar{q}q\rangle \left[\langle\bar{q}g_s\sigma Gq\rangle + \langle\bar{s}g_s\sigma Gs\rangle\right]\right\}}{6144\pi^4} \int_{y_i}^{y_f}dy\int_{z_i}^{\zeta}dz  \left(y s+2y\omega\right)\\
\notag &&- \frac{7 m_s m_c^2 \langle\bar{q}g_s\sigma Gq\rangle \left[-2 \langle\bar{q}q\rangle + \langle\bar{s}s\rangle\right]}{3072\pi^4}  \int_{y_i}^{y_f}dy\int_{z_i}^{\zeta}dz  \, \frac{1}{y}\, ,
\end{eqnarray}

\begin{eqnarray}
\notag \rho_b^0(10)&=&  \frac{m_c\langle g_s^2GG\rangle \langle\bar{q}q\rangle \left[\langle\bar{q}q\rangle + 14 \langle\bar{s}s\rangle\right]}{18432\pi^4}  \int_{y_i}^{y_f}dy\int_{z_i}^{\zeta}dz  \, \frac{z \xi}{y^2} \left(y-\frac{2}{3}\right)\\
\notag &&+ \frac{m_c \langle g_s^2GG\rangle \langle\bar{q}q\rangle \left[\langle\bar{q}q\rangle + 28 \langle\bar{s}s\rangle\right]}{36864\pi^4}  \int_{y_i}^{y_f}dy\int_{z_i}^{\zeta}dz \\
\notag &&- \frac{m_c \langle\bar{q}g_s\sigma Gq\rangle \left[24 \langle\bar{q}g_s\sigma Gq\rangle + 35 \langle\bar{s}g_s\sigma Gs\rangle\right]}{294912\pi^4}
 \int_{y_i}^{y_f}dy\int_{z_i}^{\zeta}dz  \, .
\end{eqnarray}

The $c$ type integrals:

\begin{eqnarray}
\notag \rho_c^1(9)&=&   - \frac{m_s  g_s^2  \langle\bar{q}q\rangle^2 \left[14 \langle\bar{q}q\rangle - 13 \langle\bar{s}s\rangle\right]}{124416\pi^4} \int_{y_i}^{y_f}dy  \, \zeta y\widetilde{m}_c^2+\frac{m_s \langle\bar{q}q\rangle^2 \langle\bar{s}s\rangle}{2304\pi^2} \int_{y_i}^{y_f}dy  \, \zeta y\widetilde{m}_c^2 \, ,
\end{eqnarray}

\begin{eqnarray}
\notag \rho_c^1(10)&=&  \frac{11\langle\bar{q}g_s\sigma Gq\rangle \langle\bar{s}g_s\sigma Gs\rangle}{24576\pi^4}
\int_{y_i}^{y_f}dy  \, \zeta y\widetilde{m}_c^2  + \frac{\langle g_s^2GG\rangle \langle\bar{q}q\rangle \left[\langle\bar{q}q\rangle + 14 \langle\bar{s}s\rangle\right]}{110592\pi^4} \int_{y_i}^{y_f}dy  \, \zeta y\widetilde{m}_c^2 \\
\notag &&- \frac{m_s m_c \langle\bar{q}g_s\sigma Gq\rangle \left[-39 \langle\bar{q}g_s\sigma Gq\rangle + 14 \langle\bar{s}g_s\sigma Gs\rangle\right]}{36864\pi^4} \int_{y_i}^{y_f}dy  \, \zeta \left(1+\frac{\widetilde{m}_c^2}{2 T^2}\right)\\
\notag &&- \frac{m_s m_c \langle g_s^2GG\rangle \langle\bar{q}q\rangle \langle\bar{s}s\rangle}{36864\pi^4} \int_{y_i}^{y_f}dy  \, \frac{\zeta}{y}  - \frac{m_s m_c \langle g_s^2GG\rangle \langle\bar{q}q\rangle \left[-26 \langle\bar{q}q\rangle + 7 \langle\bar{s}s\rangle\right]}{110592\pi^4}
\int_{y_i}^{y_f}dy  \, \zeta \left(1+\frac{\widetilde{m}_c^2}{2 T^2}\right)\\
\notag &&- \frac{7 m_s m_c \langle\bar{q}g_s\sigma Gq\rangle \left[3 \langle\bar{q}g_s\sigma Gq\rangle - \langle\bar{s}g_s\sigma Gs\rangle\right]}{18432\pi^4} \int_{y_i}^{y_f}dy  \, \frac{\zeta}{y} \\
\notag && +\frac{ m_s m_c \langle\bar{q}g_s\sigma Gq\rangle \langle\bar{s}g_s\sigma Gs\rangle}{18432\pi^4} \int_{y_i}^{y_f}dy  \, \zeta \left(1+\frac{\widetilde{m}_c^2}{2 T^2}\right) \, ,
\end{eqnarray}

\begin{eqnarray}
\notag \rho_c^1(11)&=&  \frac{13 m_c \langle\bar{q}q\rangle \left[2 \langle\bar{q}g_s\sigma Gq\rangle \langle\bar{s}s\rangle + \langle\bar{q}q\rangle \langle\bar{s}g_s\sigma Gs\rangle\right]}{2304\pi^2} \int_{y_i}^{y_f}dy  \, \zeta \left(1+\frac{\widetilde{m}_c^2}{2 T^2}\right)\\
\notag &&+  \frac{7 m_c \langle\bar{q}g_s\sigma Gq\rangle  g_s^2 \langle\bar{s}s\rangle^2}{124416\pi^4} \int_{y_i}^{y_f}dy  \, \zeta \left(1+\frac{\widetilde{m}_c^2}{2 T^2}\right)\\
\notag &&- \frac{ m_c \langle\bar{q}q\rangle \left[28 \langle\bar{q}g_s\sigma Gq\rangle \langle\bar{s}s\rangle + \langle\bar{q}q\rangle \langle\bar{s}g_s\sigma Gs\rangle\right]}{4608\pi^2} \int_{y_i}^{y_f}dy  \, \frac{\zeta}{y} \\
\notag &&+ \frac{m_c  g_s^2 \langle\bar{q}q\rangle^2 \left[7 \langle\bar{q}g_s\sigma Gq\rangle + \langle\bar{s}g_s\sigma Gs\rangle\right]}{124416\pi^4} \int_{y_i}^{y_f}dy  \, \zeta \left(1+\frac{\widetilde{m}_c^2}{2 T^2}\right)\\
\notag &&- \frac{m_s \langle\bar{q}q\rangle \left[3 \langle\bar{q}g_s\sigma Gq\rangle \langle\bar{s}s\rangle + \langle\bar{q}q\rangle \langle\bar{s}g_s\sigma Gs\rangle\right]}{6912\pi^2} \int_{y_i}^{y_f}dy  \, \zeta y \left(3+\frac{\widetilde{m}_c^4}{2  T^4}+\frac{2 \widetilde{m}_c^2}{ T^2}\right)\\
\notag &&+ \frac{m_s  g_s^2 \langle\bar{q}q\rangle^2 \left[21 \langle\bar{q}g_s\sigma Gq\rangle - 13 \langle\bar{s}g_s\sigma Gs\rangle\right]}{373248\pi^4} \int_{y_i}^{y_f}dy  \, \zeta y \left(3+\frac{\widetilde{m}_c^4}{2  T^4}+\frac{2 \widetilde{m}_c^2}{ T^2}\right)\\
\notag &&+ \frac{ m_s \langle\bar{q}q\rangle \langle\bar{q}g_s\sigma Gq\rangle \langle\bar{s}s\rangle}{2304\pi^2} \int_{y_i}^{y_f}dy  \, \zeta \left(1+\frac{\widetilde{m}_c^2}{2 T^2}\right)\, ,
\end{eqnarray}

\begin{eqnarray}
\notag \rho_c^1(12)&=&  \frac{ g_s^2 \langle\bar{q}q\rangle^2 \langle\bar{s}s\rangle \left[14 \langle\bar{q}q\rangle + \langle\bar{s}s\rangle\right]}{46656\pi^2} \int_{y_i}^{y_f}dy  \, \zeta y \left(3+\frac{\widetilde{m}_c^4}{2  T^4}+\frac{2 \widetilde{m}_c^2}{ T^2}\right)\\
\notag && -\frac{7 m_s m_c  g_s^2 \langle\bar{s}s\rangle \langle\bar{q}q\rangle^3}{93312\pi^2} \int_{y_i}^{y_f}dy  \, \zeta \left(\frac{\widetilde{m}_c^4}{ T^6}+\frac{2 \widetilde{m}_c^2}{ T^4}+\frac{2}{ T^2}\right) \, ,
\end{eqnarray}

\begin{eqnarray}
\notag \rho_c^1(13)&=&   -\frac{13 m_c \langle\bar{q}g_s\sigma Gq\rangle \left[\langle\bar{q}g_s\sigma Gq\rangle \langle\bar{s}s\rangle + 2 \langle\bar{q}q\rangle \langle\bar{s}g_s\sigma Gs\rangle\right]}{18432\pi^2} \int_{y_i}^{y_f}dy  \, \zeta \left(\frac{2}{ T^2}+\frac{2\widetilde{m}_c^2}{ T^4}+\frac{\widetilde{m}_c^4}{ T^6}
\right)\\
\notag &&+ \frac{13 m_c^3 \langle g_s^2GG\rangle \langle\bar{q}q\rangle^2 \langle\bar{s}s\rangle}{82944\pi^2T^4} \int_{y_i}^{y_f}dy  \, \frac{1}{y^2}  +  \frac{13 m_c\langle g_s^2GG\rangle \langle\bar{q}q\rangle^2 \langle\bar{s}s\rangle }{41472\pi^2} \int_{y_i}^{y_f}dy  \, \frac{\zeta}{y^2} \left(\frac{\widetilde{m}_c^2 y}{2 T^4}-\frac{1}{ T^2}+\frac{y}{2 T^2}\right) \\
\notag && -\frac{13 m_c \langle g_s^2GG\rangle \langle\bar{q}q\rangle^2 \langle\bar{s}s\rangle}{55296\pi^2} \int_{y_i}^{y_f}dy  \, \zeta\left(\frac{\widetilde{m}_c^4}{ T^6}+\frac{2 \widetilde{m}_c^2}{ T^4}+\frac{2}{ T^2}\right)\\
\notag &&+ \frac{m_c \langle\bar{q}g_s\sigma Gq\rangle \left[14 \langle\bar{q}g_s\sigma Gq\rangle \langle\bar{s}s\rangle + 15 \langle\bar{q}q\rangle \langle\bar{s}g_s\sigma Gs\rangle\right]}{9216\pi^2} \int_{y_i}^{y_f}dy  \, \frac{\zeta}{y} \left(\frac{\widetilde{m}_c^2}{T^4}+\frac{1}{ T^2}\right)\\
\notag &&+ \frac{m_s \langle\bar{q}g_s\sigma Gq\rangle \left[3 \langle\bar{q}g_s\sigma Gq\rangle \langle\bar{s}s\rangle + 4 \langle\bar{q}q\rangle \langle\bar{s}g_s\sigma Gs\rangle\right]}{110592\pi^2} \int_{y_i}^{y_f}dy  \, \zeta y \left(\frac{6 }{ T^2}+\frac{6 \widetilde{m}_c^2 }{ T^4}+\frac{3 \widetilde{m}_c^4}{ T^6}+\frac{ \widetilde{m}_c^6}{ T^8}\right)\\
\notag && -\frac{m_s m_c^2 \langle g_s^2GG\rangle \langle\bar{q}q\rangle^2 \langle\bar{s}s\rangle}{82944\pi^2} \int_{y_i}^{y_f}dy  \, \frac{\zeta}{y^2} \frac{\widetilde{m}_c^2}{ T^6}\\
\notag &&+ \frac{m_s \langle g_s^2GG\rangle \langle\bar{q}q\rangle^2 \langle\bar{s}s\rangle}{165888\pi^2} \int_{y_i}^{y_f}dy  \, \zeta y  \left(\frac{ \widetilde{m}_c^6}{ T^8}+\frac{3 \widetilde{m}_c^4}{ T^6}+\frac{6 \widetilde{m}_c^2}{ T^4}+\frac{6}{ T^2}\right) \\
\notag &&- \frac{m_s \langle\bar{q}g_s\sigma Gq\rangle \left[3 \langle\bar{q}g_s\sigma Gq\rangle \langle\bar{s}s\rangle + 2 \langle\bar{q}q\rangle \langle\bar{s}g_s\sigma Gs\rangle\right]}{27648\pi^2}  \int_{y_i}^{y_f}dy  \, \zeta \left(\frac{\widetilde{m}_c^4}{2 T^6}+\frac{\widetilde{m}_c^2}{ T^4}+\frac{1}{ T^2}\right)\, ,
\end{eqnarray}

\begin{eqnarray}
\notag \rho_c^0(9)&=& \frac{m_s m_c g_s^2  \langle\bar{q}q\rangle^2 \left[-14 \langle\bar{q}q\rangle + 13 \langle\bar{s}s\rangle\right]}{124416\pi^4} \int_{y_i}^{y_f}dy  \, y\widetilde{m}_c^2+\frac{m_s m_c \langle\bar{q}q\rangle^2 \langle\bar{s}s\rangle}{2304\pi^2} \int_{y_i}^{y_f}dy  \, y\widetilde{m}_c^2 \, ,
\end{eqnarray}

\begin{eqnarray}
\notag \rho_c^0(10)&=&  \frac{11 m_c \langle\bar{q}g_s\sigma Gq\rangle \langle\bar{s}g_s\sigma Gs\rangle}{24576\pi^4} \int_{y_i}^{y_f}dy  \, y\widetilde{m}_c^2 \\
\notag &&+ \frac{m_c\langle g_s^2GG\rangle \langle\bar{q}q\rangle \left[\langle\bar{q}q\rangle + 14 \langle\bar{s}s\rangle\right]}{110592\pi^4} \int_{y_i}^{y_f}dy  \, y\widetilde{m}_c^2 \\
\notag &&+ \frac{m_s m_c^2 \langle\bar{q}g_s\sigma Gq\rangle \left[39 \langle\bar{q}g_s\sigma Gq\rangle - 14 \langle\bar{s}g_s\sigma Gs\rangle\right]}{73728\pi^4} \int_{y_i}^{y_f}dy   \left(1+\frac{\widetilde{m}_c^2}{T^2}\right)\\
\notag && -\frac{m_s m_c^2 \langle g_s^2GG\rangle \langle\bar{q}q\rangle \langle\bar{s}s\rangle}{36864 \pi^4} \int_{y_i}^{y_f}dy   \, \frac{1}{y} \\
\notag &&+ \frac{ m_s m_c^2\langle g_s^2GG\rangle \langle\bar{q}q\rangle \left[26 \langle\bar{q}q\rangle - 7 \langle\bar{s}s\rangle\right]}{221184\pi^4} \int_{y_i}^{y_f}dy   \left(1+\frac{\widetilde{m}_c^2}{ T^2}\right)\\
\notag && -\frac{7 m_s m_c^2 \langle\bar{q}g_s\sigma Gq\rangle \left[3 \langle\bar{q}g_s\sigma Gq\rangle - \langle\bar{s}g_s\sigma Gs\rangle\right]}{18432\pi^4}
 \int_{y_i}^{y_f}dy   \, \frac{1}{y} \\
\notag &&+  \frac{m_s m_c^2 \langle\bar{q}g_s\sigma Gq\rangle \langle\bar{s}g_s\sigma Gs\rangle}{36864\pi^4} \int_{y_i}^{y_f}dy   \left(1+\frac{\widetilde{m}_c^2}{ T^2}\right)\, ,
\end{eqnarray}

\begin{eqnarray}
\notag \rho_c^0(11)&=&  \frac{13 m_c^2 \langle\bar{q}q\rangle \left[2 \langle\bar{q}g_s\sigma Gq\rangle \langle\bar{s}s\rangle + \langle\bar{q}q\rangle \langle\bar{s}g_s\sigma Gs\rangle\right]}{4608\pi^2} \int_{y_i}^{y_f}dy  \left(1+\frac{\widetilde{m}_c^2}{T^2}\right)\\
\notag &&+ \frac{m_c^2\left\{7 \langle\bar{q}g_s\sigma Gq\rangle  g_s^2 \langle\bar{s}s\rangle^2 + g_s^2 \langle\bar{q}q\rangle^2 \left[7 \langle\bar{q}g_s\sigma Gq\rangle + \langle\bar{s}g_s\sigma Gs\rangle\right]\right\}}{248832\pi^4} \int_{y_i}^{y_f}dy  \left(1+\frac{\widetilde{m}_c^2}{T^2}\right)\\
\notag && -\frac{m_c^2 \langle\bar{q}q\rangle \left[28 \langle\bar{q}g_s\sigma Gq\rangle \langle\bar{s}s\rangle + \langle\bar{q}q\rangle \langle\bar{s}g_s\sigma Gs\rangle\right]}{4608\pi^2} \int_{y_i}^{y_f}dy  \, \frac{1}{y}\\
\notag &&- \frac{m_s m_c \langle\bar{q}q\rangle \left[3 \langle\bar{q}g_s\sigma Gq\rangle \langle\bar{s}s\rangle + \langle\bar{q}q\rangle \langle\bar{s}g_s\sigma Gs\rangle\right]}{6912\pi^2}
 \int_{y_i}^{y_f}dy  \, y \left(1+\frac{\widetilde{m}_c^4}{2 T^4}+\frac{\widetilde{m}_c^2}{T^2}\right)\\
\notag &&+ \frac{m_s m_c  g_s^2 \langle\bar{q}q\rangle^2 \left[21 \langle\bar{q}g_s\sigma Gq\rangle - 13 \langle\bar{s}g_s\sigma Gs\rangle\right]}{373248\pi^4} \int_{y_i}^{y_f}dy  \, y \left(1+\frac{\widetilde{m}_c^4}{2 T^4}+\frac{\widetilde{m}_c^2}{ T^2}\right)\\
\notag &&+  \frac{ m_s m_c \langle\bar{q}q\rangle \langle\bar{q}g_s\sigma Gq\rangle \langle\bar{s}s\rangle }{4608\pi^2} \int_{y_i}^{y_f}dy  \left(1+\frac{\widetilde{m}_c^2}{T^2}\right)\, ,
\end{eqnarray}

\begin{eqnarray}
\notag \rho_c^0(12)&=&  \frac{m_c  g_s^2 \langle\bar{q}q\rangle^2 \langle\bar{s}s\rangle \left[14 \langle\bar{q}q\rangle + \langle\bar{s}s\rangle\right]}{46656\pi^2} \int_{y_i}^{y_f}dy  \, y \left(1+\frac{\widetilde{m}_c^4}{2 T^4}+\frac{\widetilde{m}_c^2}{T^2}\right)\\
\notag &&-  \frac{7 m_s m_c^2  g_s^2 \langle\bar{q}q\rangle^3 \langle\bar{s}s\rangle}{93312\pi^2} \int_{y_i}^{y_f}dy  \frac{\widetilde{m}_c^4}{ T^6}\, ,
\end{eqnarray}

\begin{eqnarray}
\notag \rho_c^0(13)&=&  -\frac{13 m_c^2 \langle\bar{q}g_s\sigma Gq\rangle \left[\langle\bar{q}g_s\sigma Gq\rangle \langle\bar{s}s\rangle + 2 \langle\bar{q}q\rangle \langle\bar{s}g_s\sigma Gs\rangle\right]}{18432\pi^2} \int_{y_i}^{y_f}dy  \, \frac{\widetilde{m}_c^4}{ T^6}\\
\notag &&+ \frac{13 m_c^2\langle g_s^2GG\rangle \langle\bar{q}q\rangle^2 \langle\bar{s}s\rangle}{20736\pi^2} \int_{y_i}^{y_f}dy  \, \frac{1}{y^2} \left(\frac{1}{T^2}-\frac{\widetilde{m}_c^2 y}{2 T^4}\right)\\
\notag &&-  \frac{13 m_c^2 \langle g_s^2GG\rangle \langle\bar{q}q\rangle^2 \langle\bar{s}s\rangle}{55296\pi^2} \int_{y_i}^{y_f}dy  \, \frac{\widetilde{m}_c^4}{ T^6}\\
\notag &&+ \frac{m_c^2 \langle\bar{q}g_s\sigma Gq\rangle \left[14 \langle\bar{q}g_s\sigma Gq\rangle \langle\bar{s}s\rangle + 15 \langle\bar{q}q\rangle \langle\bar{s}g_s\sigma Gs\rangle\right]}{9216\pi^2} \int_{y_i}^{y_f}dy  \, \frac{\widetilde{m}_c^2}{y T^4} \\
\notag &&+ \frac{m_s m_c \langle\bar{q}g_s\sigma Gq\rangle \left[3 \langle\bar{q}g_s\sigma Gq\rangle \langle\bar{s}s\rangle + 4 \langle\bar{q}q\rangle \langle\bar{s}g_s\sigma Gs\rangle\right]}{110592\pi^2} \int_{y_i}^{y_f}dy  \, \frac{y\widetilde{m}_c^6}{ T^8}\\
\notag && -\frac{m_s m_c \langle\bar{q}g_s\sigma Gq\rangle \left[3 \langle\bar{q}g_s\sigma Gq\rangle \langle\bar{s}s\rangle + 2 \langle\bar{q}q\rangle \langle\bar{s}g_s\sigma Gs\rangle\right]}{55296\pi^2} \int_{y_i}^{y_f}dy  \, \frac{\widetilde{m}_c^4}{T^6}\\
\notag &&+ \frac{m_s m_c \langle g_s^2GG\rangle \langle\bar{q}q\rangle^2 \langle\bar{s}s\rangle}{165888\pi^2} \int_{y_i}^{y_f}dy   \left(\frac{3\zeta\widetilde{m}_c^2}{y^2 T^4}+\frac{y \widetilde{m}_c^6}{ T^8}\right)\\
\notag &&- \frac{ m_s m_c^3\langle g_s^2GG\rangle \langle\bar{q}q\rangle^2 \langle\bar{s}s\rangle}{165888\pi^2} \int_{y_i}^{y_f}dy  \left(\frac{\zeta}{y^3}+\frac{1}{y^2}\right) \left(\frac{\widetilde{m}_c^2}{ T^6}-\frac{1}{ T^4}\right)\, .
\end{eqnarray}

The $d$ type integrals:

\begin{eqnarray}
\notag \rho_d^1(7)&=&  - \frac{m_s m_c^2 \langle g_s^2GG\rangle \left[-28 \langle\bar{q}q\rangle + 13 \langle\bar{s}s\rangle\right]}{294912\pi^6} \int_{y_i}^{y_f}dy\int_{z_i}^{\zeta}dz  \, \frac{z \xi^2\overline{m}_c^2}{y^2} \, ,
\end{eqnarray}

\begin{eqnarray}
\notag \rho_d^1(10) &=&\frac{m_c^2\langle g_s^2GG\rangle \langle\bar{q}q\rangle \left[\langle\bar{q}q\rangle + 14 \langle\bar{s}s\rangle\right]}{27648\pi^4} \int_{y_i}^{y_f}dy\int_{z_i}^{\zeta}dz  \, \frac{z \xi}{y^2} \left(1+\frac{\overline{m}_c^2}{2 T^2}\right)\\
\notag &&+ \frac{\langle g_s^2GG\rangle \langle\bar{q}q\rangle \left[\langle\bar{q}q\rangle + 20 \langle\bar{s}s\rangle\right]}{73728\pi^4} \int_{y_i}^{y_f}dy\int_{z_i}^{\zeta}dz  \, z\overline{m}_c^2 \\
\notag &&- \frac{\langle\bar{q}g_s\sigma Gq\rangle \left[24 \langle\bar{q}g_s\sigma Gq\rangle + 35 \langle\bar{s}g_s\sigma Gs\rangle\right]}{589824\pi^4} \int_{y_i}^{y_f}dy\int_{z_i}^{\zeta}dz  \, z\overline{m}_c^2\\
\notag &&+ \frac{m_s m_c \langle g_s^2GG\rangle \langle\bar{q}q\rangle \left[-13 \langle\bar{q}q\rangle + 7 \langle\bar{s}s\rangle\right]}{55296\pi^4} \int_{y_i}^{y_f}dy\int_{z_i}^{\zeta}dz  \, \frac{z}{y^2} \left(\frac{\overline{m}_c^2 y}{2 T^2}-\zeta\right)\\
\notag &&+ \frac{ m_s m_c^3\langle g_s^2GG\rangle \langle\bar{q}q\rangle \left[-13 \langle\bar{q}q\rangle + 7 \langle\bar{s}s\rangle\right]}{110592\pi^4T^2} \int_{y_i}^{y_f}dy\int_{z_i}^{\zeta}dz  \, \frac{1}{y^2 } \, ,
\end{eqnarray}

\begin{eqnarray}
\notag \rho_d^0(7)&=& \frac{m_s m_c^3 \langle g_s^2GG\rangle \left[28 \langle\bar{q}q\rangle - 13 \langle\bar{s}s\rangle\right]}{589824\pi^6} \int_{y_i}^{y_f}dy\int_{z_i}^{\zeta}dz  \left(\frac{\xi^2}{y^2}+\frac{z \xi^2}{y^3}\right) \,\overline{m}_c^2\, ,
\end{eqnarray}

\begin{eqnarray}
\notag \rho_d^0(10)&=&  \frac{m_c^3\langle g_s^2GG\rangle \langle\bar{q}q\rangle \left[\langle\bar{q}q\rangle + 14 \langle\bar{s}s\rangle\right]}{110592\pi^4} \int_{y_i}^{y_f}dy\int_{z_i}^{\zeta}dz  \, \frac{\xi}{y^2} \left(1+\frac{\overline{m}_c^2}{ T^2}\right)\\
\notag &&- \frac{m_c\langle g_s^2GG\rangle \langle\bar{q}q\rangle \left[\langle\bar{q}q\rangle + 14 \langle\bar{s}s\rangle\right]}{27648\pi^4} \int_{y_i}^{y_f}dy\int_{z_i}^{\zeta}dz  \, \frac{z \xi}{y^2} \left(\frac{\overline{m}_c^2}{2}-\frac{\overline{m}_c^4 y}{4 T^2}-\overline{m}_c^2 y\right)\\
\notag &&+ \frac{m_c\langle g_s^2GG\rangle \langle\bar{q}q\rangle \left[\langle\bar{q}q\rangle + 28 \langle\bar{s}s\rangle\right]}{73728\pi^4} \int_{y_i}^{y_f}dy\int_{z_i}^{\zeta}dz  \, \overline{m}_c^2 \\
\notag &&- \frac{m_c \langle\bar{q}g_s\sigma Gq\rangle \left[24 \langle\bar{q}g_s\sigma Gq\rangle + 35 \langle\bar{s}g_s\sigma Gs\rangle\right]}{589824\pi^4} \int_{y_i}^{y_f}dy\int_{z_i}^{\zeta}dz   \, \overline{m}_c^2\\
\notag &&+ \frac{m_s m_c^2 \langle g_s^2GG\rangle \langle\bar{q}q\rangle \left[13 \langle\bar{q}q\rangle - 7 \langle\bar{s}s\rangle\right]}{27648\pi^4} \int_{y_i}^{y_f}dy\int_{z_i}^{\zeta}dz  \, \frac{1}{y^2} \left(1-\frac{\overline{m}_c^2 y}{2 T^2}-\frac{y}{2}\right).
\notag
\end{eqnarray}

\section*{Acknowledgements}
This  work is supported by National Natural Science Foundation, Grant Number  12175068.

\end{document}